\documentclass{article}

\usepackage{microtype}
\usepackage{graphicx}
\usepackage{subcaption}
\usepackage{booktabs}

\usepackage{hyperref}

\makeatletter
\@namedef{ver@algorithmic.sty}{}
\makeatother

\usepackage[preprint]{icml2026} 
\usepackage{amsmath,amssymb,mathtools,amsthm}

\usepackage{url}
\usepackage{multirow}

\usepackage{algorithm}
\usepackage{algpseudocode}

\usepackage[capitalize,noabbrev]{cleveref}

\usepackage[textsize=tiny]{todonotes}

\icmltitlerunning{SurfelSoup: Learned Point Cloud Geometry Compression With a Probablistic SurfelTree Representation}

\begin{document}

\twocolumn[
  % \icmltitle{SurfelSoup: Probabilistic pSurfelTree for \\
  % Learned Point Cloud Geometry Compression}
  % \icmltitle{SurfelSoup: Surface-Based Point Cloud Geometry Compression \\ with Learned pSurfelTree}
  \icmltitle{SurfelSoup: Learned Point Cloud Geometry Compression \\ With a Probablistic SurfelTree Representation}



  \icmlsetsymbol{equal}{*}

  \begin{icmlauthorlist}
    \icmlauthor{Tingyu Fan}{nyu}
    \icmlauthor{Yueyu Hu}{nyu}
    \icmlauthor{Ran Gong}{nyu}
    \icmlauthor{Yao Wang}{nyu}
  \end{icmlauthorlist}

    \icmlaffiliation{nyu}{
    Department of Electrical and Computer Engineering, 
    New York University Tandon School of Engineering, 
    Brooklyn, NY, USA
    }

    \icmlcorrespondingauthor{Tingyu Fan}{tf2387@nyu.edu}

  \icmlkeywords{Point Cloud Compression, Learned Compression}

  \vskip 0.3in
]



\printAffiliationsAndNotice{}  % no special notice (required even if empty)

\begin{abstract}
This paper presents SurfelSoup, an end-to-end learned surface-based framework for point cloud geometry compression, with surface-structured primitives for representation. It proposes a probabilistic  surface representation, pSurfel, which models local point occupancies using a bounded generalized Gaussian distribution. In addition, the pSurfels are organized into an octree-like hierarchy, pSurfelTree, with a Tree Decision module that adaptively terminates the tree subdivision  for rate-distortion optimal Surfel granularity selection. 
This formulation avoids redundant point-wise compression in smooth regions and produces compact yet smooth surface reconstructions. Experimental results under the MPEG common test condition show consistent gain on geometry compression over voxel-based baselines and MPEG standard G-PCC-GesTM-TriSoup, while providing visually superior reconstructions with smooth and coherent surface structures.
\end{abstract}

\section{Introduction}
% Point clouds have emerged as a crucial 3D data representation in the field of immersive media, autonomous driving, and virtual/augmented reality~\citep{ICLR2025_804f1fac, ICLR2025_8e86fd92}.  To this end, the Moving Picture Experts Group (MPEG) established two standards for point cloud compression (PCC): video-based PCC (V-PCC)~\citep{schwarz2018emerging} and geometry-based PCC (G-PCC)~\citep{zhang2024standard}. Beyond these standards, numerous learning-based methods have emerged~\citep{quach2020improved,wang2021lossy,wang2021multiscale,wang2023dynamic,wang2024versatile1,que2021voxelcontext,ijcai2022p126,you2025reno}. One key challenge is the uneven geometry structure of the points. Compared with 2D images/videos, points in a point cloud are irregularly spread over a relatively small portion of the 3D space, making the design of the context model and the reconstruction of the point cloud  difficult.

Point clouds, which represent the 3D structures using colored points, have emerged as a crucial 3D data representation in the field of immersive media, autonomous driving, and virtual/augmented reality~\citep{ICLR2025_804f1fac, ICLR2025_8e86fd92}, where compression is necessary to reduce the storage and transmission costs. To this end, the Moving Picture Experts Group (MPEG) established two standards for point cloud compression (PCC): video-based PCC (V-PCC)~\citep{schwarz2018emerging} and geometry-based PCC (G-PCC)~\citep{zhang2024standard}. Beyond these standards, numerous learning-based methods have emerged~\citep{quach2020improved,wang2021lossy,wang2021multiscale,wang2023dynamic,wang2024versatile1,que2021voxelcontext,ijcai2022p126,you2025reno}. Despite these advancements, the fact that points are unevenly distributed in the 3D space poses significant challenges to the design of  efficient PCC schemes.

The main difficulty of point cloud compression lies in establishing an expressive and compression-friendly 3D representation. G-PCC-Octree uses the octree structure, which transfers the non-uniform points into a tree. To exploit the surface prior, G-PCC-TriSoup represents a point cloud as a set of triangles, significantly outperforming G-PCC-Octree
on dense point clouds.  Meanwhile, several learned point cloud compression methods~\citep{wang2021lossy,wang2022sparse,wang2024versatile1,wang2025unipcgc} have been proposed to leverage sparse convolution~\citep{choy20194d} on the voxelized point clouds. Although these learned approaches improve compression efficiency, they remain fundamentally tied to voxelized representations and octree-based structures, with redundant voxel divisions in smooth regions ignoring the underlying surface structure, leading to inherently less efficient compression.
In addition, voxel-based methods often struggle to preserve continuous surfaces and tend to produce artifacts such as gaps at low bit rates.
%limited flexibility for surface structures.

% Inspired by G-PCC-TriSoup, this paper aims at learned surface-based representation for point cloud geometry compression.  Textured Surfel Octree (TeSO) \citep{hu2025teso} employs cube-bounded surfels~\citep{pfister2000surfels} organized in an octree to represent 3D scene, each parameterized by a normal vector, an octree bounding box, a center, and a radius. 
% A TeSO can be directly constructed from a given point cloud, by checking whether the occupied points in an octree node  is sufficiently approximated by a surfel under a chosen distortion criterion. However, such
% construction does not optimize the rate vs. distortion trade-off, because it does not consider the compression of the surfel parameters when deciding whether to split a node further. 
% This paper aims to train a deep-learning model that can directly generate the TeSO geometry (and consequently decoded point coordinates) that is rate-distortion optimal.
%However, using the TeSO representation as is makes the model training challenging: 
%(1) Surfels have no thickness, (2) Geometry distortion is difficult to measure when approximating points with a surfel. (3) Surfels' generation does not consider rate-distortion trade off.
% However,   TeSO suffers from: 
% %TeSO suffers from two key limitations: 
% (1) Surfels have no thickness; (2) Geometry distortion is difficult to measure when approximating points with a surfel; (3) TeSO construction does not optimize  rate-distortion trade-off.

 \begin{figure}[tbp]
  \centering
    \includegraphics[width=0.99\linewidth]{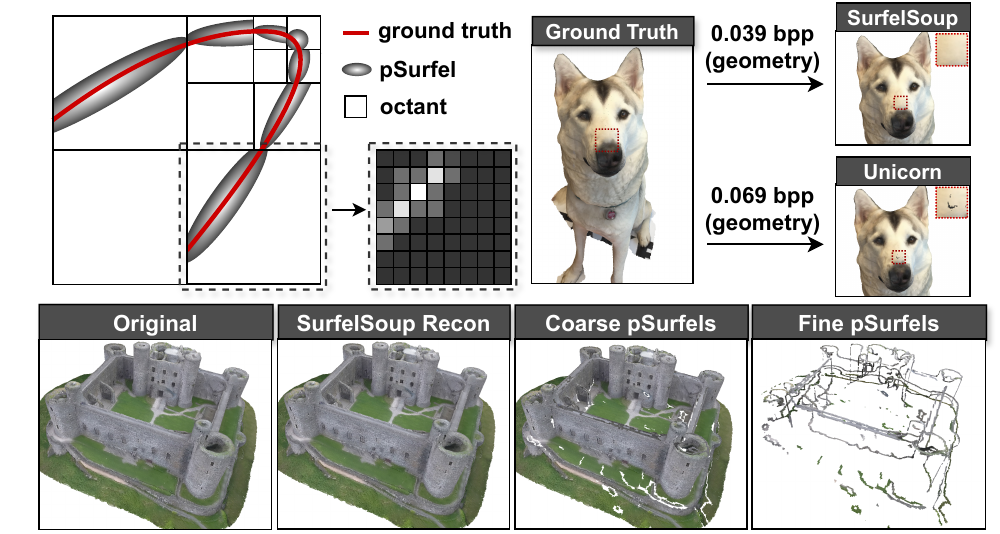}
   \vspace{-2mm}\caption{An illustration of SurfelSoup. Top Left: a toy 2D example for the pSurfelTree structure. Top Right: Example reconstruction comparison of a 3D Husky. Bottom: Example of pSurfelTree's multi-granularity reconstruction.} 
   \label{fig:intro}
   \vspace{-3mm}
\end{figure}

% Note that point-to-plane distance is not applicable because it cannot constrain the surfel radius. Using point-to-point  distortion  on the other hand requires sampling of the surfel, which makes the distortion non-differentiable.

% To this end, we propose SurfelSoup, the first end-to-end learned surface-based point cloud compression framework, which models geometry directly as a composition of probabilistic surfaces termed pSurfels.  pSurfel models voxel occupancy as a bounded 3D generalized Gaussian distribution. Unlike TeSO’s deterministic, zero-thickness surfels, our probabilistic formulation is continuous and differentiable, allowing the end-to-end optimization.
To this end, we break from the voxel-based paradigm and introduce SurfelSoup. To the best of our knowledge, this is the first end-to-end surface-based framework for point cloud compression, with surface primitives for reconstruction. SurfelSoup models a point cloud as a composition of probabilistic surfaces, termed pSurfels, organized in an octree-like structure called pSurfelTree (Fig.~\ref{fig:intro}). Each pSurfel describes  voxel occupancy likelihoods within an octree node using a differentiable bounded 3D generalized Gaussian distribution, enabling end-to-end optimization.
% With the probabilistic voxel occupancy, binary cross entropy provides a differentiable distortion measurement. 
% By minimizing this distortion, the trained model will generate a 3D GG function that has a small variance in one direction (approximating the surfel normal), close to a 2D GG function on the tangent plane, when the underlying points in the node fall on a surfel.
%We further organize the pSurfels  in a pSurfelTree (Fig.~\ref{fig:intro}), with
% Our model generates features at occupied nodes in a fine-to-coarse manner following the octree structure. Starting at a selected coarse level, it quantizes and codes the features at all the occupied nodes. At each node, a decision module determines if the node should terminate as a surfel. If yes, it further generates the surfel parameters from the quantized features. Otherwise, it estimates the likelihood that each children node is occupied, based on the feature as well as the context  generated from the features of the sibling nodes that have been decoded. This process repeats at each finer level using the features interpolated from those decoded at the coarser level, until a chosen finest level is reached.   
A Tree Decision module adaptively decides whether to terminate a tree node with a pSurfel, or to further split it into children nodes,
%, enabling a  more refined surface description, 
based on the desired rate-distortion trade-off. Experiments show that, compared with voxel baselines, SurfelSoup achieves greater coding efficiency and qualitatively superior reconstructions with smooth and coherent surface structures.
%Specifically,
%enabling the adaptive assignments of pSurfels at different scales (octree depths) under  unified end-to-end rate–distortion optimization.
% generates the likelihood of a tree node should terminate with a pSurfel. We formulate the expected total distortion and total rate, respectively,  based on the estimated surfel likelihoods at all nodes across all levels.  This formulation enables us to train all the modules together using a loss that is a weighted sum of the distortion and rate.
%We evaluate  SurfelSoup under the MPEG-specified common test conditions (CTC). Our experiment results show that SurfelSoup outperforms the previous state-of-the-art learning-based method Unicorn~\citep{wang2024versatile1}, as well as G-PCC-Trisoup.
%as well as other voxel-based methods for dense point cloud geometry compression. 
Our main contributions include:
\begin{itemize}
    \item {\bf The first end-to-end learned surface-based point cloud geometry compression framework, SurfelSoup}, leveraging a probabilistic surface representation, termed {\bf pSurfel}. This representation models the voxel occupancies by a bounded 3D generalized Gaussian function, which induces a differentiable likelihood over voxel occupancies for end-to-end optimization.
    %thereby eliminating the need to encode redundant voxel subdivisions in smooth regions.
    \item {\bf A pSurfelTree hierarchy for adaptive surfel granularity assignment}. pSurfelTree makes rate-distortion optimal assignment of pSurfels at different granularities. This is achieved through the proposed Tree Decision module, and a corresponding formulation of the  expected total rate and distortion.
    %adaptively terminates nodes at different scales (octree levels) as pSurfels under the rate-distortion trade-off.   
  %  adaptive selection and compression of pSurfels at different scales (octree depths).

    \item {\bf Validations according to the MPEG common test condition (CTC) for AI-PCC}. Our experimental results show that SurfelSoup outperforms the previous state-of-the-art voxel-based method Unicorn~\citep{wang2024versatile1} and other voxel-based approaches, as well as MPEG G-PCC-TriSoup, under the Common Test Condition. Models trained using human datasets according to the MPEG CTC further show strong generalization to object and scene point clouds.
    % The pSurfel representation further enables fast and high-quality free-viewpoint rendering.

\end{itemize}

\section{Related Work}
\subsection{Rule-based Point Cloud Compression}
The Moving Picture Experts Group (MPEG) specifies video-based PCC (V-PCC) and geometry-based PCC (G-PCC). V-PCC projects 3D point clouds onto 2D. G-PCC encodes the geometry directly in 3D using regularized data structures. G-PCC-Octree represents the geometry as an occupancy octree, while G-PCC-TriSoup approximates points under each octree node with  triangles at a chosen octree level.
%surfaces with an unordered set of triangles (``triangle soup”). 
%For dense point clouds with explicit surface structure, 
%For 3D scenes that are primarily composed of surfaces,
TriSoup has been shown to outperform G-PCC-Octree~\citep{quach2020improved} for dense point clouds, as it  encodes surfaces with far fewer parameters than the voxel occupancies.

\subsection{Learning-based Point Cloud Compression}
Early learning-based point cloud compression methods borrow the idea of end-to-end entropy models~\citep{balle2017end,balle2018variational,minnen2018joint} and adopt 3D convolutions to process volumetric inputs~\citep{quach2020improved,wang2021lossy,nguyen2021learning}. However, the excessive computational complexity of dense 3D convolutions requires pre-partitioning of the input~\citep{quach2020improved,wang2021lossy}, resulting in long encoding/decoding times and degraded efficiency. To address this, PCGCv2~\citep{wang2021multiscale} first introduces 3D sparse convolutions~\citep{choy20194d} into point cloud compression, greatly reducing computational cost. Building upon this, SparsePCGC~\citep{wang2022sparse} proposes a unified $N$-stage Sparse Octree Probability Aggregation (SOPA) module for both lossless and lossy compression, which auto-regressively models octant occupancy distribution in a fixed Morton order. ViewPCGC~\citep{zheng2024viewpcgc} further introduces view information into point cloud compression.
%Due to the flexibility of arbitrarily combining lossless and lossy SOPA, as well as  the efficiency of auto-regressive modeling, SparsePCGC achieves significant gains over PCGCv2. 
More recently, UniPCGC~\citep{wang2025unipcgc} introduces a Variable Rate and Complexity Module that enables adaptive rate control, further improving over SparsePCGC.  Unicorn~\citep{wang2024versatile1} proposes a unified architecture for geometry and attribute compression, which incorporates a Neighborhood Point Attention (NPA) module for geometry compression. There are also works that focus on LiDAR point clouds~\citep{huang2020octsqueeze,biswas2020muscle,fu2022octattention,you2025reno}, exploiting sparsity and range-image structures. These designs are effective for LiDAR, but do not generalize well to non-LiDAR point clouds.

 \begin{figure*}[tbp]
  \centering
    \includegraphics[width=0.9\linewidth]{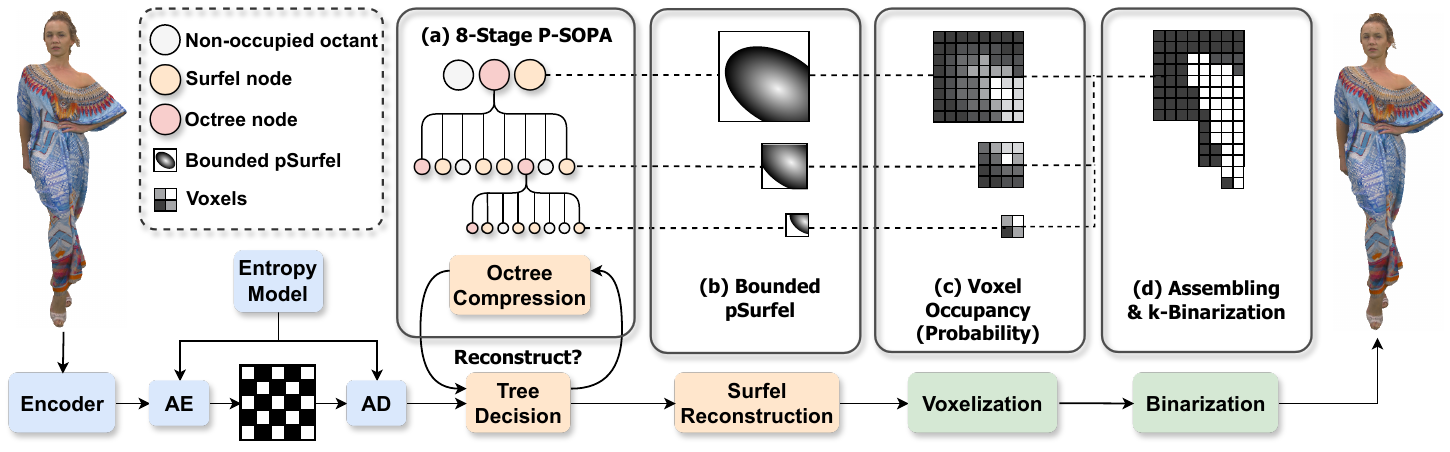}
   \vspace{-2mm}\caption{The overall architecture of SurfelSoup. Encoder and Entropy Model correspond to Sec.~\ref{chap:encoder}. Surfel Reconstruction, Voxelization and Binarization correspond to Sec.~\ref{chap:surfel_recon}. Decision corresponds to Sec.~\ref{chap:decision}. Octree Compression corresponds to Sec.~\ref{chap:SOPA}.} 
   \label{fig:overall}
   \vspace{-3mm}
\end{figure*}

\subsection{3D Representations}

% Recent advances on surface representation for 3D data can be divided as implicit and explicit. 

% Key points: place our work in the context of 3D Gaussians, surface representation and point cloud compression
% The problem we are trying to solve is different from 3D gaussians splatting.
% We are not re-inventing surfels. We use the idea of explicit surface representation to efficiently compress point cloud.
% One most related work is TeSO.

% Surface representation is a fundamental research problem in computer vision and computer graphics.
% There have been works exploring implicit \cite{xxx} and explicit surface representation.

Representing 3D surfaces for reconstruction, processing, and rendering has been a fundamental research problem. To this end, both implicit~\citep{park2019deepsdf,fridovich2022plenoxels,mildenhall2021nerf,chibane2020neural} and explicit surface representations~\citep{pfister2000surfels,dai2024high,gao2023surfelnerf,aliev2020neural} have been developed. Notably, surfel~\citep{pfister2000surfels} proposes to represent a surface with isolated surface elements without storing the connectivity, hence achieving efficient rendering and processing flexibility. Later works introduce the surface splatting~\citep{zwicker2001surface,kobbelt2004survey} techniques and non-square shapes~\citep{botsch2005high} to address the visibility issues in rendering. By designing the primitives with differentiability, surfels and more generally 3D Gaussians have been shown to facilitate 3D reconstruction from multi-view images and novel view rendering~\cite{kerbl20233d,dai2024high}.

% Our work is inspired by~\citep{pfister2000surfels}, but we extend the surfel primitives to differentiable pSurfels to facilitate end-to-end point cloud compression.  
% Our work is orthogonal to 3D Gaussian splatting, where the goal is to reconstruct 3D structures from multi-view images, instead of point cloud compression (Appendix~\ref{appendix:compareGS}). 
% Recent continuous 3D representations such as neural implicit surfaces and 3D Gaussian splatting are designed for reconstruction and rendering via per-scene optimization. In contrast, SurfelSoup is explicitly formulated for learned point cloud geometry compression under an entropy-constrained rate–distortion objective.

However, the above representations, including neural implicit surfaces and 3D Gaussian splatting, are primarily designed for 3D reconstruction and rendering, where the objective is to fit 3D structures to a given scene. In contrast, SurfelSoup is explicitly formulated for learned point cloud geometry compression under a rate--distortion objective. A systematic comparison with these representations is provided in Appendix~\ref{appendix:compareGS}. Our work is inspired by~\citep{pfister2000surfels}, but extends the surfel primitives to differentiable pSurfels for end-to-end point cloud compression.

Our work is most related to TeSO~\citep{hu2025teso}, where the 3D scene is described by cube-bounded surfels organized on an octree, with the geometry of each surfel  parameterized by a normal vector, an octree bounding box, a center, and a radius. However, the TeSO is constructed  based on a hand-crafted criterion then compressed without the joint optimization, thus leading to rate-distortion suboptimality.

\section{Methodology}
\subsection{Overview}\label{chap:overall}
The architecture of SurfelSoup is illustrated in Figure~\ref{fig:overall}. We provide a notation table in Appendix~\ref{appendix:notation}. The input point cloud ${\bf p}$ is represented as a sparse tensor ${\bf p}=\left[{\bf p}_C,{\bf p}_A\right]$, where ${\bf p}_C$ denotes the 3D coordinates and ${\bf p}_A$ the associated attributes. For geometry compression, ${\bf p}_A=\mathbf{1}$ to indicate occupancy. The  encoder module (Sec.~\ref{chap:encoder}) maps ${\bf p}$ into latent ${\bf f}^L$ via sparse convolution, where the superscript ``$L$'' denotes $L$ stages of octree-based downsampling. ${\bf f}^L$ is quantized, arithmetically coded and decoded (AE/AD) to yield $\hat{\bf f}^L$. For each non-empty node ${\bf u}_i^L$ with latent $\hat{\mathbf{f}}^L_i$, the Tree Decision module (Sec.~\ref{chap:decision}) selects between:
\begin{enumerate}
\item reconstructing a pSurfel via the surfel reconstruction module (Sec.~\ref{chap:surfel_recon}), or

\item compressing the occupancy information of the eight child octants of the current node using the octree compression module (Sec.~\ref{chap:SOPA}).
\end{enumerate}

The above process is then recursively applied on non-empty octants generated by Decision 2. The recursive process yields a tree-structured representation, termed the pSurfelTree, where a  node is a leaf if and only if it reconstructs a pSurfel. At octree layer $l \in [0,L]$, we denote by ${\bf s}^l$ the set of nodes following Decision 1 (surfel nodes), and ${\bf o}^l$ the set of nodes following Decision 2 (octree nodes). Finally, each surfel node ${\bf s}^l_i$ is converted into a local point set with assigned occupancy likelihoods based on the decoded pSurfel parameters via the voxelization module,
%and binarization modules, 
and all such sets are assembled and reconstructed into a point cloud through the binarization module.
 % top $K$ points are chosen to form the reconstructed point cloud $\hat{\bf p}$.

\subsection{Encoder \& Entropy Coding} {\label{chap:encoder}}
The encoder module maps the input point cloud ${\bf p}$ into a latent ${\bf f}^L$. It consists of $L$ downsampling stages following the design of \citep{wang2024versatile1}, implemented via sparse convolutions. The latent representation is decomposed as ${\bf f}^L=\left[{\bf f}^L_C, {\bf f}^L_A\right]$, where the geometric coordinates ${\bf f}^L_C$ are losslessly compressed using the G-PCC-Octree, while the attributes ${\bf f}^L_A$ (containing the latent features) are quantized and entropy coded using the hyperprior entropy model~\citep{balle2018variational}. The reconstructed attribute latent $\hat{\bf f}^L$ is then upsampled to each octree level $l$ via sparse transposed convolutions (See Appendix~\ref{appendix:ns}), yielding $\hat{\bf f}^l,l=1,2,\ldots, L-1$ for subsequent pSurfelTree construction.

 \begin{figure*}[tbp]
  \centering
    \includegraphics[width=0.8\linewidth]{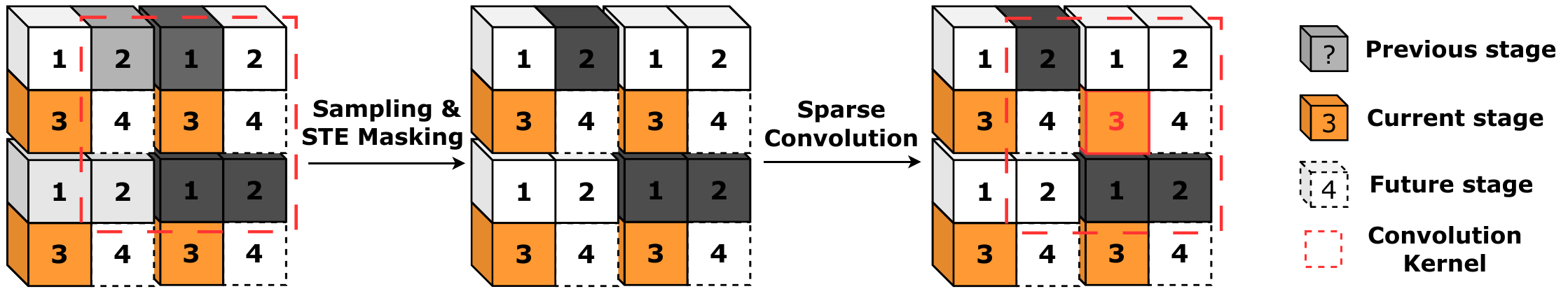}
    \caption{A 2D example to explain P-SOPA. The octants labeled 1 and 2 have been coded, with gray  intensities indicating the existence probabilities (the darker the higher probability). Octants labeled 3 are being coded, and octants labeled 4 are yet to be coded and unseen. In the  middle figure, some octants labeled 1 and 2 were randomly masked out (set as white). Real P-SOPA is done on 3D.} 
   \vspace{-2mm}
   \label{fig:psopa}
\end{figure*}

\subsection{pSurfel Representation}{\label{chap:surfel_recon}}
We model local geometry using a learnable surface primitive, termed pSurfel, which defines a continuous occupancy likelihood field within a spatial cell. Concretely, a pSurfel is parameterized as a bounded 3D generalized Gaussian:
\begin{equation}
\begin{aligned}
    P({\bf x}) &= \exp\!\left(
        -\tfrac{1}{2}\, r({\bf x},{\bf \Sigma})^{2\beta}
    \right), \\
    r({\bf x},{\bf \Sigma}) &= 
        \sqrt{({\bf x}-{\bf \mu})^T {\bf \Sigma}^{-1} ({\bf x}-{\bf \mu})}.
\end{aligned}
\label{eq_gg}
\end{equation}

where $\bf x\in\mathbb{R}^3$ denotes any location  in 3D space. 
%The center $\mu$'s likelihood is $1$, because the formulation of the pSurfelTree ensures at least one point  present inside each surfel node.

 Following 3D Gaussian splatting~\citep{kerbl20233d}, each pSurfel associated with node ${\bf s}_i^l$ is parameterized by 11 learnable variables: mean ${\bf \mu}_i^l\in \mathbb{R}^3$, standard deviation ${\sigma}_i^l\in \mathbb{R}^3$, quaternion ${\bf q}_i^l\in \mathbb{R}^4$ 
% (which have 3 parameters)
, and shape coefficient $\beta_i^l\in\mathbb{R}$.

Note that the 3D GG function reduces to a 2D function describing a planar surfel when one of its rotated axes has a very small variance. This axis is the normal vector of the surfel,  and the other two axes define  two orthogonal axes on the surfel. The $\sigma_i$ along each plane axis defines the spread of the surfel. 
The parameter $\beta$ controls the sharpness of the drop off beyond the respective $\sigma$ in each axis.
% appropriate because we want the predicted occupancy at a voxel to be close to either 1 or 0, depending on if the  voxel is   on the surfel or not.
The GG function reduces to the  Gaussian function when $\beta=2$, which has slower drop-off.
% (less accurate) predicted occupancy.

\vspace{-2mm}\begin{algorithm}[t]
\caption{pSurfelTree Construction}
\label{alg:gsurfeltree}
\begin{algorithmic}[1]
\Require Latent feature $\hat{\mathbf f}_i^l$, octants $\mathcal{O}_i^l$
\Function{ConstructTree}{$\hat{\mathbf f}_i^l$}
  \State $\tilde{p}_i^l, \tilde{q}_i^l \gets \Call{DecisionModule}{\hat{\mathbf f}_i^l}$
  \State $\mathcal{D}(\mathbf s_i^l) \gets \Call{SurfelRecon}{\hat{\mathbf f}_i^l}$
  \State $\mathcal B(\mathbf o_i^l),\, \hat{\mathbf f}_{i,[0:7]}^l 
         \gets \Call{P\mbox{-}SOPA}{\hat{\mathbf f}_i^l}$
  \For{$j = 0$ \textbf{to} $7$}
    \If{$\mathcal O^l_{i,j} = 1$ \textbf{and} (train \textbf{or} $\tilde q_i^l < \epsilon$)}
      \State \Call{ConstructTree}{$\hat{\mathbf f}_{i,j}^l$}
    \EndIf
  \EndFor
\EndFunction
\end{algorithmic}
\end{algorithm}
 
\vspace{-1mm}\paragraph{Bounding Box.} 
%To enable pSurfels to exploit the boundaries of the bounding boxes to construct  complex geometries, we restrict the influence of each pSurfel to the bounding box defined by the octree, {\it i.e.} the coverage region of ${\bf s}_i^l$ with side length $2^l$ , as shown in Fig.~\ref{fig:overall} (b).
We restrict the influence of each pSurfel to the bounding box defined by the octree, {\it i.e.} the coverage region of ${\bf s}_i^l$ with side length $2^l$ , as shown in Fig.~\ref{fig:overall} (b). Such bounded pSurfels at different granularities can represent complex surface geometries.

\vspace{-2mm}\paragraph{Voxelization and Distortion.} %The continuous field in Eq.~\ref{eq_gg} induces the likelihood of voxel occupancies. 
Eq.~\ref{eq_gg} defines a continuous occupancy probability field over 3D space.
For each voxel ${\bf v}_{i,k}^l$ within the bounding box, we evaluate its occupancy probability $P({\bf v}_{i,k}^l)$.
% based on the pSurfel parameters according to Eq.~\ref{eq_gg}. 
 The distortion incurred by classifying ${\bf u}_i^l$ as a surfel node ${\bf s}_i^l$ is  defined as the negative log-likelihood of the ground truth occupancies:
\begin{equation}\label{eq:distortion}
\begin{aligned}
\mathcal{D}({\bf u}_i^l)
= &-\!\!\sum_{{\bf v}_{i,k}^l \in \mathcal{V}_i^l}
\Big[
     Q({\bf v}_{i,k}^l)\,\log P({\bf v}_{i,k}^l) \\
    & + (1 - Q({\bf v}_{i,k}^l))\,\log\!\left(1 - P({\bf v}_{i,k}^l)\right)
\Big],
\end{aligned}
\end{equation}
where $Q({\bf v}_{i,k}^l)$ 
% $\mathcal{O}_{i,j}^l$
denotes the ground truth occupancy of ${\bf v}_{i,k}^l$, $\mathcal{V}_i^l$ denotes the set of voxels within the bounding box. 

\vspace{-2mm}\paragraph{Binarization.} At the inference time,  the binarization module  converts the occupancy probability into binary occupancy codes for point cloud reconstruction. Specifically, all $\mathcal{V}_i^l$ are assembled into a global set $\mathcal{V}$, among which the top $\rho N$ voxels with the highest occupancy probabilities are selected as the reconstructed point cloud, where $N$ denotes the number of points in the original point cloud, and $\rho$ is a user-defined scaling factor (set to $1$ in our experiment).

\vspace{-2mm}\begin{algorithm}[t]
\caption{P-SOPA}
\label{alg:psopa}
\begin{algorithmic}[1]
\Require Latent feature $\hat{\mathbf f}_i^l$, octants $\mathcal{O}_i^l$
\Function{P\mbox{-}SOPA}{$\hat{\mathbf f}_i^l$}
  \For{$j = 0$ \textbf{to} $7$}
    \State $\mathcal P^l_{[:j-1]} \gets \{\tilde{p}^l_{i',j'} \mid \forall\, i', j' \in [0{:}j-1]\}$ 
    \State $m^l_{[:j-1]} \sim \mathrm{Bernoulli}\!\left(\mathcal P^l_{[:j-1]}\right)$
    \State $\bar m^l_{[:j-1]} \gets \Call{STE}{m^l_{[:j-1]}}$
    \State $[\theta, \hat{\mathbf f}]_{i,j}^l \gets 
           \Call{SOPA}{\hat{\mathbf f}_i^{l},\, \mathcal O^l_{[:j-1]},\, \bar m^l_{[:j-1]}}$
  \EndFor
  \State $\mathcal B(\mathbf o_i^l) \gets 
         \Call{ArithmCoder}{\theta_i^l,\, \mathcal O_i^l}$
  \State \Return $\mathcal B(\mathbf o_i^l),\, \hat{\mathbf f}_{i,[0:7]}^l$
\EndFunction
\end{algorithmic}
\end{algorithm}

\subsection{pSurfelTree Decision} \label{chap:decision}
%The decision module reconstructs the point cloud using pSurfels of different sizes (Note that the size of a pSurfel depends on both its variance and the size of bounding box defined by octree depth $l\in\left[0, 3\right]$). For a node with latent $\hat{\bf f}_i^l$, 

%\begin{enumerate}
 %   \item if the point geometry within the node can be well represented by a single pSurfel, the model classifies it as a surfel node ${\bf s}_i^l$ and further predicts the pSurfel parameters , thereby avoiding the additional bits required to encode deeper octree structures. 

%\item If the point geometry is more complex, the node is classified as an octree node ${\bf o}_i^l$, its octant occupancy is encoded (see Sec.~\ref{chap:SOPA}).
%\end{enumerate}

 \begin{figure*}[tbp]
  \centering
    \includegraphics[width=0.92\linewidth]{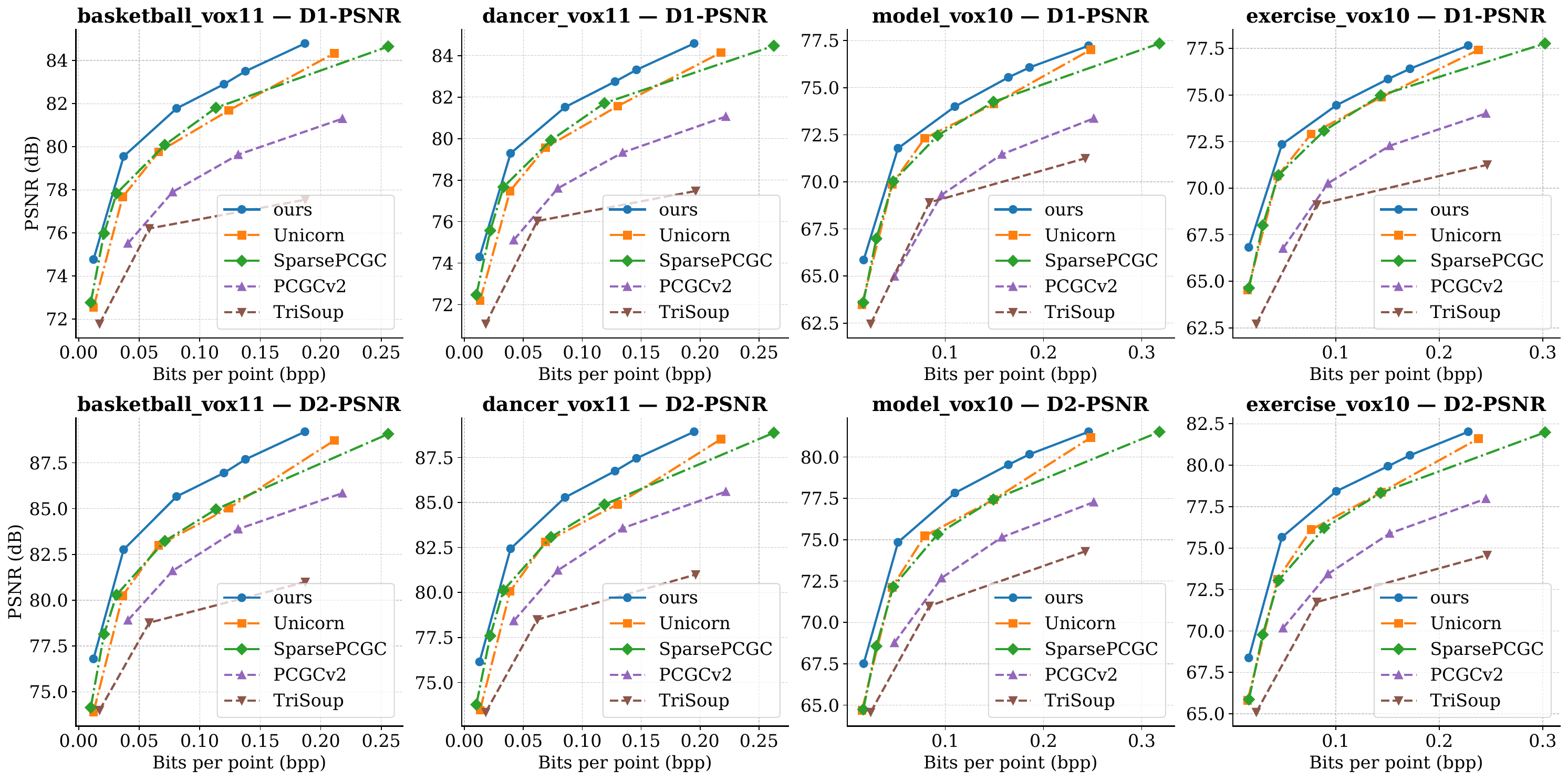}
   \caption{The comparison between SurfelSoup (ours) and baselines on Owlii. } 
   \label{fig:d1}
\end{figure*}

To adaptively assign pSurfels at different tree levels, %To adaptively assign the optimal granularity of constructing a pSurfel, 
the Tree Decision module governs whether a tree node ${\bf u}_i^l$ is further subdivided or terminates as a pSurfel.
To enable the joint training of the Tree Decision with other modules, we replace the binary decisions with differentiable probabilistic assignments during training.
However, directly computing the {\bf marginal probability} that a node ${\bf u}_i^l$ is an octree node (denoted as $\tilde{p}_i^l$) or a surfel node ($\tilde{q}_i^l$) is intractable, since a node exists only if its parent is an octree node requiring further splitting. Therefore, the decision module generates a {\bf conditional probability} $p_i^l$ that  the node ${\bf u}_i^l$ is an octree node  from its latent $\hat{\bf f}_i^l$, conditioned on either 1) its parent is also an octree node with probability $\tilde{p}_{\pi(i,l)}^{l+1}$; or 2) it is a top-level node ($l=L$). The  probabilities $\tilde{p}_i^l$ and $\tilde{q}_i^l$ are therefore recursively computed as:
\begin{equation}\label{eq:poct}
\begin{aligned}
   \tilde{p}_i^l &= p_i^l\,\tilde{p}_{\pi(i,l)}^{l+1}, 
   &\quad \tilde{p}_i^L &= p_i^L, \\
   \tilde{q}_i^l &= (1 - p_i^l)\,\tilde{p}_{\pi(i,l)}^{l+1},
   &\quad \tilde{q}_i^L &= 1 - p_i^L .
\end{aligned}
\end{equation}

where $\pi(i,l)$ denotes the index of the parent  of ${\bf u}_i^l$. 

To make the decisions differentiable while  binary during evaluation, we adopt the Gumbel–Sigmoid reparameterization~\citep{jang2017categorical}, providing a continuous sampling relaxation.

During inference, nodes with $\tilde{q}_i^l \ge 1-\epsilon$ are 
%one\text{-}hot 
classified as surfel nodes $\mathbf{s}_i^l$ and the recursion terminates. 
%We set $\epsilon=?$ in our experiments. 
During training, however, no hard termination is applied; each node is represented in a probabilistic manner, allowing all branches to remain active and differentiable. The overall tree construction procedure is summarized in Alg.~\ref{alg:gsurfeltree}.

\subsection{Octree Node Occupancy Compression: P-SOPA}\label{chap:SOPA}
We model an octree node's octant occupancies using a local autoregressive probability model: 8-stage SOPA~\citep{wang2021multiscale}. Define the octants of an octree node ${\bf o}_i^l$ as $\mathcal{O}_i^l=\left\{\mathcal{O}_{i,j}^l\mid j\in\left\{0,1,\cdots,7\right\}\right\}$, where $\mathcal{O}_{i,j}^l\in\left\{0,1\right\}$ denotes the $j$-th octant in Morton order. For each octree layer $l$, octants with the same index $j$, {\it i.e.} $\mathcal{O}_{\left[j\right]}^l=\left\{\mathcal{O}_{i,j}^l|\forall i\right\}$, are grouped together and coded in parallel. 
% Specifically,  a neural network $\Phi_j^l $ with parameter $\phi_j^l$ first generates the interpolated latents for group $j$ octants from $\mathcal{O}_{[0:j-1]}^l$ and $ \hat{\bf f}^{l}=\bigcup_i \hat{\bf f}^{l}$ through sparse convolution:
Specifically, SOPA estimates the conditional probability $\theta_{[j]}^l$ of $\mathcal{O}_{\left[j\right]}^l$ given the already decoded octants $\mathcal{O}_{[:j-1]}^l$ and latent $\hat{\bf f}^l$:
\begin{equation}
\theta_{[j]}^l=\Psi_j^l\left(\hat{\bf f}_{\left[j\right]}^l, \hat{\bf f}^{l};\psi_{j}^l\right),\quad
\hat{\bf f}_{\left[j\right]}^l=\Phi_j^l\!\left(\mathcal{O}_{[:j-1]}^l, \hat{\bf f}^{l}; \phi_{j}^l\right),
 \end{equation}
where $\Psi_j^l/\Phi_j^l(\cdot)$ are sparse convolution networks for probability estimation/latent reconstruction, parameterized by $\psi_j^l/\phi_j^l$. $\hat{\bf f}_{\left[j\right]
}=\left\{\hat{\bf f}_{i,j}^l|\forall i\right\}$ is the union of group $j$'s reconstructed latents, which integrates the information of $\hat{\bf f}^l$ and already decoded context $\mathcal{O}_{\left[j\right]}^l$.

When ${\bf u}_i^l$ is classified as an octree node ${\bf o}_i^l$, the bits required for compressing the octant occupancy information is  the negative log-likelihood of octant occupancies:
\begin{equation}\label{eq:bits}
    \mathcal{B}({\bf u}_i^l) =
    -\sum_{j=0}^7 \Big[ 
        \mathcal{O}_{i,j}^l \log_2 \theta_{i,j}^l
        \\ +(1-\mathcal{O}_{i,j}^l)\log_2 \big(1-\theta_{i,j}^l\big)
    \Big].
\end{equation}
 \begin{figure*}[tbp]
  \centering
    \includegraphics[width=0.9\linewidth]{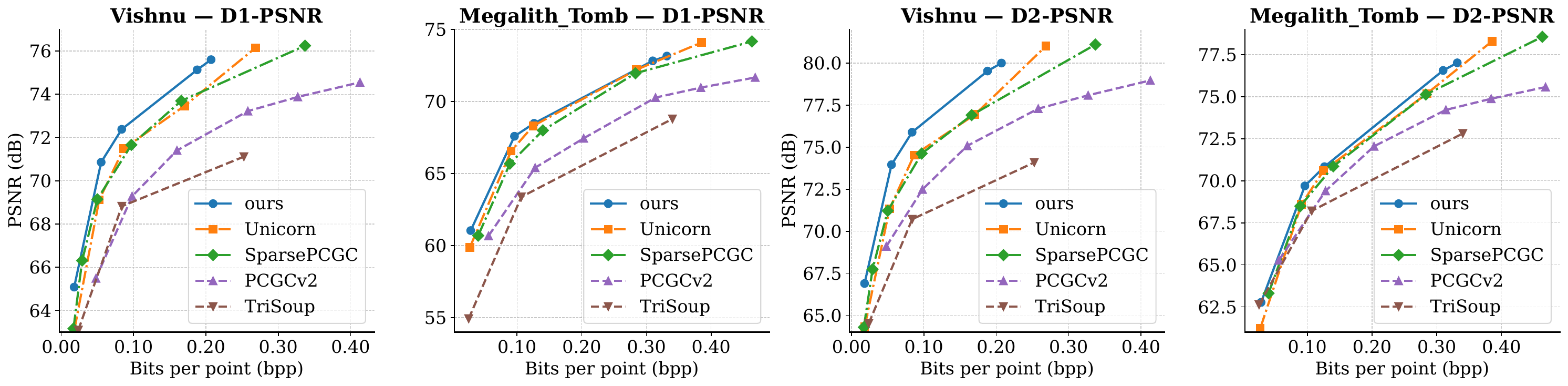}
   \caption{The comparison between SurfelSoup (ours) and baselines on RWTT dataset.} 
   \label{fig:d1-d2-rwtt}
\end{figure*}

% \renewcommand{\arraystretch}{1.4} % taller rows
% \setlength{\tabcolsep}{8pt}       % horizontal padding
% \begin{table*}[t]
% \footnotesize
% \centering
% \vspace{-2mm}
% \caption{BD-rate (\%) comparison of our method against different baselines. Negative values indicate bitrate savings over the baseline.}
% \label{tab:bdrate}
% \resizebox{\textwidth}{!}{
% \begin{tabular}{|l|c|c|c|c|c|c|c|c|c|c|c|c|}
% \hline
% \multirow{2}{*}{Baseline}
%  & \multicolumn{2}{|c|}{Basketball}
%  & \multicolumn{2}{|c|}{Dancer}
%  & \multicolumn{2}{|c|}{Model}
%  & \multicolumn{2}{|c|}{Exercise}
%  & \multicolumn{2}{|c|}{Megalith Tomb}
%  & \multicolumn{2}{|c|}{Vishnu} \\
% \cline{2-13}
%  & D1 & D2 & D1 & D2 & D1 & D2 & D1 & D2 & D1 & D2 & D1 & D2 \\
% \hline
% Unicorn      & -35.81 & -38.73 & -34.08 & -36.15 & -24.50 & -28.81 & -24.17 & -28.97 & -13.13 & -13.60 & -28.27 & -32.85 \\
% SparsePCGC   & -28.87 & -34.23 & -26.41 & -31.00 & -28.18 & -33.53 & -28.32 & -34.12 & -24.03 & -14.88 & -26.09 & -33.44 \\
% PCGCv2       & -69.37 & -61.99 & -68.42 & -60.51 & -69.41 & -63.52 & -68.71 & -65.43 & -51.28 & -28.55 & -60.24 & -55.98 \\
% TriSoup      & -74.98 & -73.98 & -73.62 & -72.82 & -68.44 & -69.55 & -74.14 & -71.50 & -62.95 & -29.10 & -57.28 & -63.57 \\
% \hline
% \end{tabular}}
% \vspace{-3mm}
% \end{table*}

\renewcommand{\arraystretch}{1.25}
\setlength{\tabcolsep}{6pt}
\begin{table*}[t]
\footnotesize
\centering
\vspace{-2mm}
\caption{BD-rate (\%) comparison of our method against different baselines.
Negative values indicate bitrate savings over the baseline.}
\label{tab:bdrate}
\begin{tabular}{l|cc|cc|cc|cc|cc|cc}
\toprule
\multirow{2}{*}{Baseline}
 & \multicolumn{2}{c|}{Basketball}
 & \multicolumn{2}{c|}{Dancer}
 & \multicolumn{2}{c|}{Model}
 & \multicolumn{2}{c|}{Exercise}
 & \multicolumn{2}{c|}{Megalith Tomb}
 & \multicolumn{2}{c}{Vishnu} \\
\cmidrule(r){2-3}
\cmidrule(r){4-5}
\cmidrule(r){6-7}
\cmidrule(r){8-9}
\cmidrule(r){10-11}
\cmidrule(r){12-13}
 & D1 & D2 & D1 & D2 & D1 & D2 & D1 & D2 & D1 & D2 & D1 & D2 \\
\midrule
Unicorn    
& -35.81 & -38.73 
& -34.08 & -36.15 
& -24.50 & -28.81 
& -24.17 & -28.97 
& -13.13 & -13.60 
& -28.27 & -32.85 \\

SparsePCGC 
& -28.87 & -34.23 
& -26.41 & -31.00 
& -28.18 & -33.53 
& -28.32 & -34.12 
& -24.03 & -14.88 
& -26.09 & -33.44 \\

PCGCv2     
& -69.37 & -61.99 
& -68.42 & -60.51 
& -69.41 & -63.52 
& -68.71 & -65.43 
& -51.28 & -28.55 
& -60.24 & -55.98 \\

TriSoup    
& -74.98 & -73.98
& -73.62 & -72.82
& -68.44 & -69.55
& -74.14 & -71.50
& -62.95 & -29.10
& -57.28 & -63.57 \\
\bottomrule
\end{tabular}
\vspace{-3mm}
\end{table*}

As shown in Alg.~\ref{alg:gsurfeltree}, during training each node ${\mathbf{u}}_{i,j}^l$ is always divided with probabilities rather than prematurely terminated as a surfel node.
This creates a mismatch between the training and inferencing of the autoregressive SOPA model: naive convolutions may access contextual octants whose parent nodes would not exist at evaluation time, violating the intended conditional structure (See Appendix~\ref{app:psopa} and Fig.~\ref{fig:notations} for an example).
We therefore propose \textbf{P-SOPA}, a probabilistic extension of SOPA that explicitly accounts for node existence. As illustrated in Fig.~\ref{fig:psopa}, when coding Group $j$, we sample a Bernoulli mask $m_{[:j-1]}^l$ from context $\mathcal{O}_{[:j-1]}^l$'s existence probabilities: $\mathcal{P}_{[:j-1]}^l=\left\{\mathcal{P}_{i,j'}^l=\tilde{p}_{i}^l|\forall i;j'\in\left[:j-1\right]\right\}$, where $\tilde{p}_i^l$ is defined in Eq.~\ref{eq:poct} indicating the probability of a node terminate. In addition, to preserve gradient flow, we apply a straight-through estimator (STE)~\citep{bengio2013estimating}:
\begin{equation}
    \bar{m}_{[:j-1]}^l=\mathrm{sg}\!\big(m_{[:j-1]}^l-\mathcal{P}_{[:j-1]}^l\big)+\mathcal{P}_{[:j-1]}^l,
\end{equation}
where $\mathrm{sg}(\cdot)$ denotes the stop-gradient operator. The STE mask $\bar{m}_{[:j-1]}^l$ is then applied to SOPA's context extraction network on $\mathcal{O}_{[:j-1]}^l$. More information of P-SOPA and its implementation is shown in Appendix~\ref{app:psopa} and~\ref{appendix:ns}. The whole algorithm of P-SOPA is shown in Alg.~\ref{alg:psopa}.

\subsection{Loss Function Formulated as Expectations} 
We formulate learning as a rate–distortion optimization:
\begin{equation}
    \mathcal{L}=\lambda \cdot \mathcal{D}+\mathcal{R}_f+\mathcal{R}_o,
\end{equation}
where $\mathcal{R}_f$ denotes the bit rate of compressing latent ${\bf f}^L$ using the hyperprior entropy model~\citep{balle2018variational}.
Due to the probabilistic Tree Decision, both the distortion  $\mathcal{D}({\bf u}_i^l)$ and the octree bit rate $\mathcal{R}_o({\bf u}_i^l)$, $\forall {\bf u}_i^l$, are random variables. We therefore optimize their expectations:
\begin{equation}
   % \mathcal{D}=\frac{1}{N}\sum_{l,i}\tilde{q}_i^l \mathcal{D}({\bf s}_i^l),\quad \mathcal{R}_o=\frac{1}{N}\sum_{l,i} \tilde{p}_i^l \mathcal{B}({\bf o}_i^l),
\mathcal{D}=\frac{1}{N}\sum_{l,i}\tilde{q}_i^l \mathcal{D}({\bf u}_i^l),\quad \mathcal{R}_o=\frac{1}{N}\sum_{l,i} \tilde{p}_i^l \mathcal{B}({\bf u}_i^l),
\end{equation}
where $\mathcal{D}({\bf u}_i^l)$ is the  distortion in Eq.~\ref{eq:distortion} when ${\bf u}_i^l$ is classified as a surfel node, $\mathcal{B}({\bf u}_i^l)$ is the octree coding cost in Eq.~\ref{eq:bits} when ${\bf u}_i^l$ is classified as an octree node. The  marginal probabilities $\tilde{p}_i^l$ and $\tilde{q}_i^l$ are  determined according to Eq.~\ref{eq:poct}.
%where $\mathcal{D}({\bf u}_i^l)$ is the negative log-likelihood distortion in Eq.~\ref{eq:distortion}, $\mathcal{B}({\bf u}_i^l)$ is the octree coding cost in Eq.~\ref{eq:bits}, $\tilde{p}_i^l$ and $\tilde{q}_i^l$ denote the marginal probabilities of ${\bf u}_i^l$ being an octree and a surfel node, respectively, determined according to Eq.~\ref{eq:poct}.

% We train multiple models with different $\lambda$, targeted for different bit rates. 
% Note that the total rate also includes the bits needed to code the occupancy information up to level $l=3$ using G-PCC-Octree.

\section{Experiments}
\subsection{Datasets}
 We train our model on the training dataset that the MPEG Common Test Condition (CTC) specifies: 8i Voxelized Full Bodies (8iVFB) \citep{d20178i}. We use the five sequences in 8iVFB for training, including {\it longdress}, {\it redandblack}, {\it soldier}, {\it loot} and {\it queen}.  
 We evaluate on the CTC-specified testing dataset: Owlii \citep{keming2018owlii}. We use the four sequences in Owlii for evaluation. We put the MPEG  CTC's 12-bit sequence {\it ThaiDancer} in Appendix~\ref{sec:thai}. To evaluate the generalization ability of our model, 
we also evaluate on  the object and scene dataset: Real World Texture Things (RWTT)~\citep{maggiordomo2020real}; and the scene dataset: ScanNet~\citep{dai2017scannet} (See Appendix~\ref{appendix:scannet}).

\subsection{Experiment Setup}

% \paragraph{Training Details.} We train four models with $\lambda=0.1,0.5,1.0,1.5$. We force the network to reconstruct pSurfels at octree level $l=1$, to avoid the excessive time complexity and bit rate overhead of losslessly compressing the raw point cloud ($l=0$) using 8-stage SOPA. The initial learning rate is set as 0.0002. We train each model for 1 day. 

\paragraph{Baselines.} We compare with: Unicorn~\citep{wang2024versatile1}, SparsePCGC~\citep{wang2022sparse} and PCGCv2~\citep{wang2021multiscale} retrained on the same 8iVFB dataset as our method, which show better performance under MPEG CTC test condition compared with the original version trained on ShapeNet~\citep{chang2015shapenet}. We retrain the baselines with the same training time as our method. We also compare with MPEG G-PCC-TriSoup~\citep{zhang2024standard}, which outperforms G-PCC-Octree. We did not compare with UniPCGC~\citep{wang2025unipcgc} because its Variable Rate and Complexity Module (VRCM) is for fine-grained rate control rather than improving geometry modeling.

\begin{figure*}[tbp]
  \centering
    \includegraphics[width=0.88\linewidth]{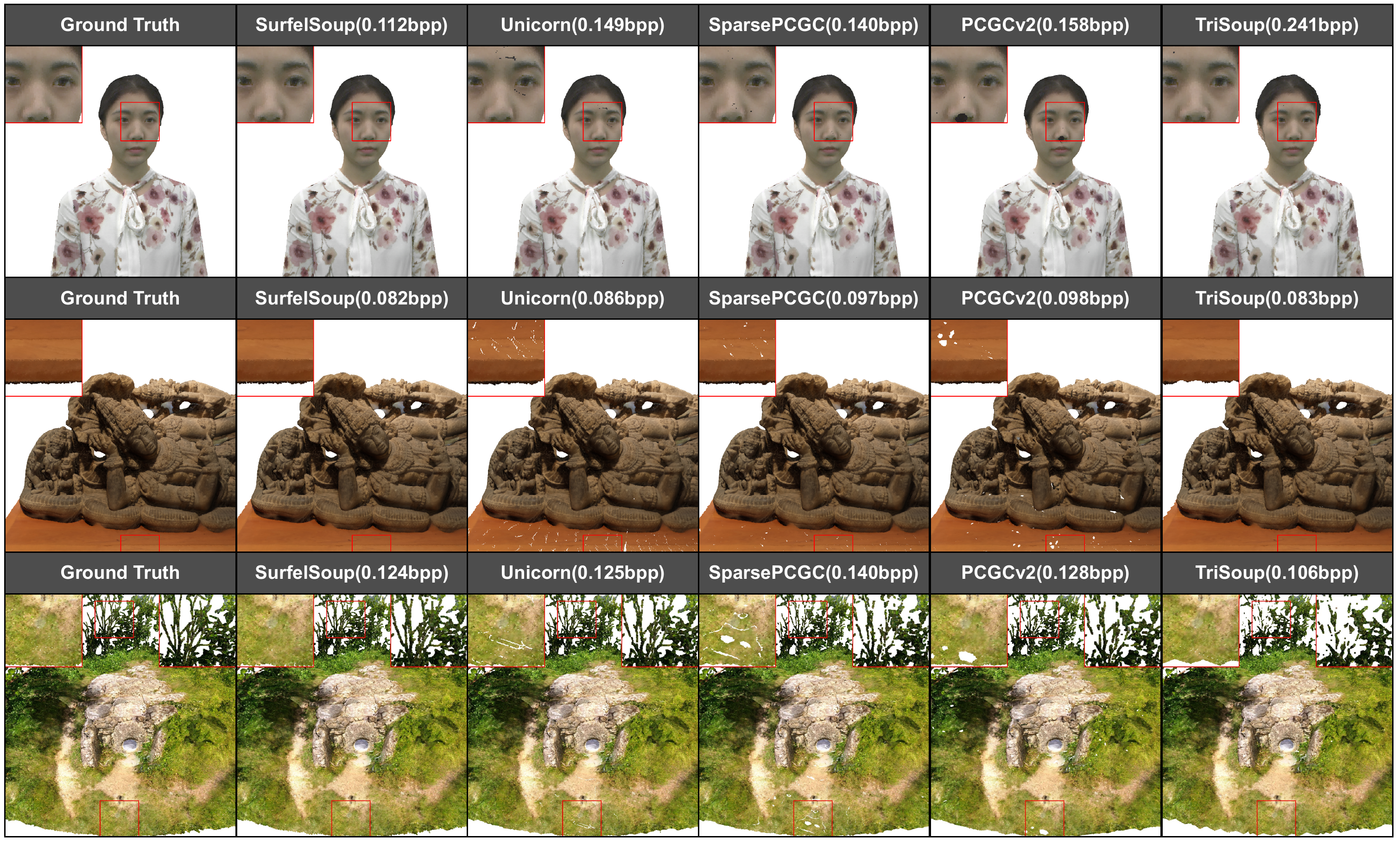}
   \caption{Visual comparison of decoded point clouds. The colors are interpolated from the original point cloud.} 
   \vspace{-4mm}
   \label{fig:visual}
\end{figure*}

% \vspace{-2mm}\paragraph{Evaluation Metrics.} We evaluate the reconstruction quality by D1- (point to point) and D2-PSNR (point to plane).

\subsection{Experimental Results}
\paragraph{Owlii.} The rate–distortion (RD) curves on Owlii are shown in Fig.\ref{fig:d1}. The BD-rate results are summarized in Tab.\ref{tab:bdrate}. Due to the pSurfel representation that compactly models smooth point cloud surfaces, our method achieves an average BD-rate of $-29.64\%$ (D1) and $-33.17\%$ (D2) over Unicorn, the current state-of-the-art method for dense point cloud geometry compression~\citep{xu2025point}. SurfelSoup also achieves consistent gain compared with other baselines. Notably, the improvements are more significant on the two 11-bit (vox11) sequences, which exhibit more smooth surfaces that are easily fitted by pSurfels.

\vspace{-2mm}\paragraph{RWTT.} To further validate the generalization ability of our model, we evaluate {\bf without finetuning} on the RWTT dataset, specifically the two sequences {\it Megalith Tomb} and {\it Vishnu}, which are also included in MPEG’s test set. The RD-curves are shown in Fig.~\ref{fig:d1-d2-rwtt}. Note that {\it Vishnu} is an object point cloud and {\it Megalith Tomb} is a scene point cloud. Our method shows great generalization ability on both non-human objects and sparse scenes. The test result on ScanNet (Appendix~\ref{appendix:scannet}) further validates our model's performance on scene point clouds.

 \vspace{-2mm}\begin{figure}[tbp]
  \centering
    \includegraphics[width=0.99\linewidth]{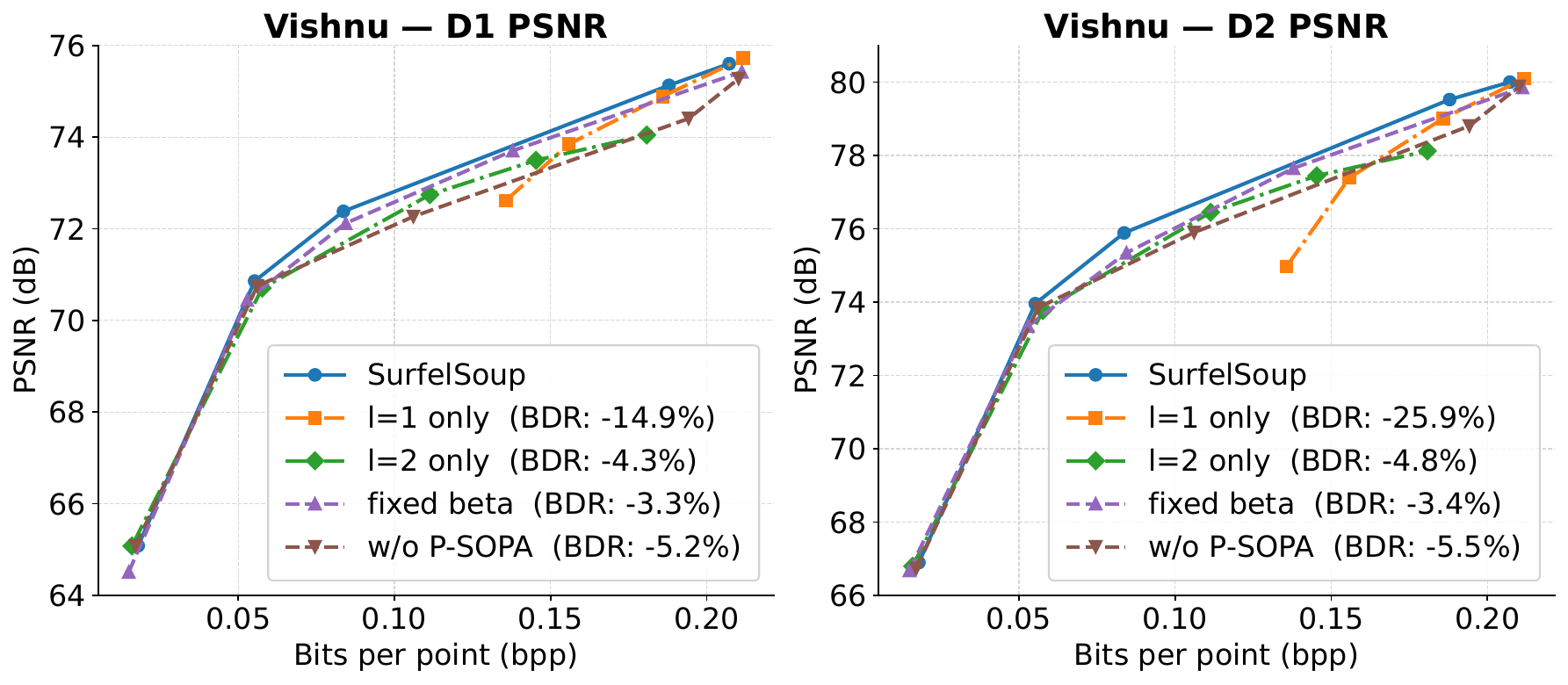}
    \vspace{-2mm}
   \caption{Ablation studies done on RWTT-Vishnu. ``$l=1$ only'' and ``$l=2$ only'': removing the pSurfelTree decision and constructs pSurfels on only one layer. ``fixed beta'': fixing the shape coefficient $\beta=2$. ``w/o P-SOPA'': replacing P-SOPA with SOPA.} 
   \vspace{-3mm}
   \label{fig:abl}
\end{figure}

\paragraph{Visual Comparison.} We compare the rendered views of compressed point clouds in Fig.~\ref{fig:visual}. We color the geometry-only reconstruction by interpolation, because human eyes are more sensitive to geometry distortion with color. Pure geometry visualization is shown in Appendix~\ref{appendix:geoonly}. Voxel-based baselines suffer from the discontinuity of voxel representations. This manifests especially under low bitrates due to {\bf error propagation}: when coarse octree layers are lossy coded, dropping a single occupied octant removes all the corresponding fine-level voxels, leading to large gaps. This issue is especially pronounced on flat surfaces, where the resulting holes and cracks become highly visible. In contrast, SurfelSoup models the whole node as a continuous surface element, which exhibits no error propagation and preserves smooth and coherent geometry even under low bit rate.

% Note that even though the geometry distortion in terms of D1 and D2 by SurfelSoup and Unicorn are similar, SufelSoup achieved superior visual quality while having lower geometry rates. More visualizations is shown in Appendix.
%By extracting the 2D surfel from each decoded pSurfel based the direction with the least variance,  and its associated texture patch, we can use the TeSO rendering method \cite{hu2025teso}, which yields further improved  rendered images.

\subsection{Ablation Studies}
\paragraph{Decision Module.} We test SurfelSoup without the decision module. As shown in Fig.~\ref{fig:abl}, forcing the pSurfel Nodes to terminate at $l=1$ only (  $2^3 $ cubes) or at $l=2$ only ($4^3 $ cubes) leads to inferior performance at low or high bitrates.

\vspace{-2mm}\paragraph{P-SOPA.}\label{sec:psopa} We test SurfelSoup trained without probabilistic masking in P-SOPA. As shown in Fig.~\ref{fig:abl}, the model has a huge performance drop in middle rate points, where the surfel nodes spread in different octree layers. Under that case, conducting SOPA on a node during training will face serious information leakage (See example in Appendix~\ref{app:psopa} and Fig.~\ref{fig:notations}). Therefore,the information leakage during training is not negligible, leading to a mismatch between training and evaluation.

\begin{figure*}[t]
  \centering
  \includegraphics[height=0.22\textheight]{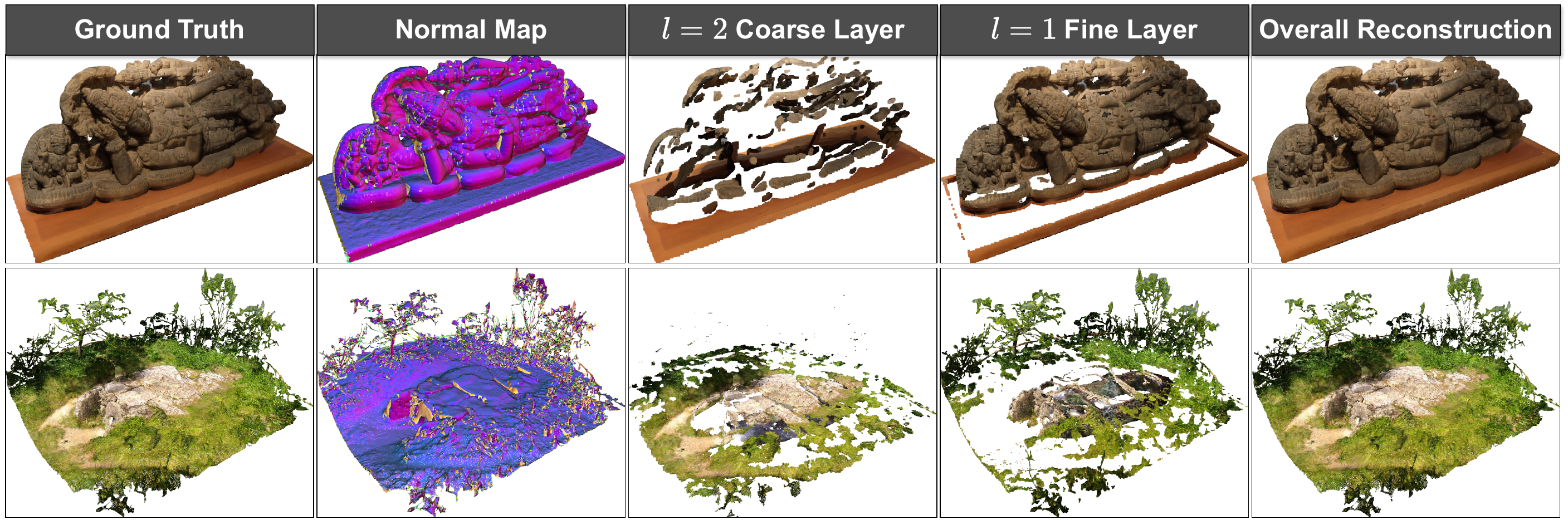}
  \caption{Visualization of different layers of decoded point clouds. 
  It is noted that coarse layers capture flat surfaces (base, ground), while fine layers capture detailed non-surface structures (Buddha statue, trees).}
  \label{fig:layers}
\end{figure*}

\begin{figure*}[t]
  \centering
  \includegraphics[height=0.25\textheight]{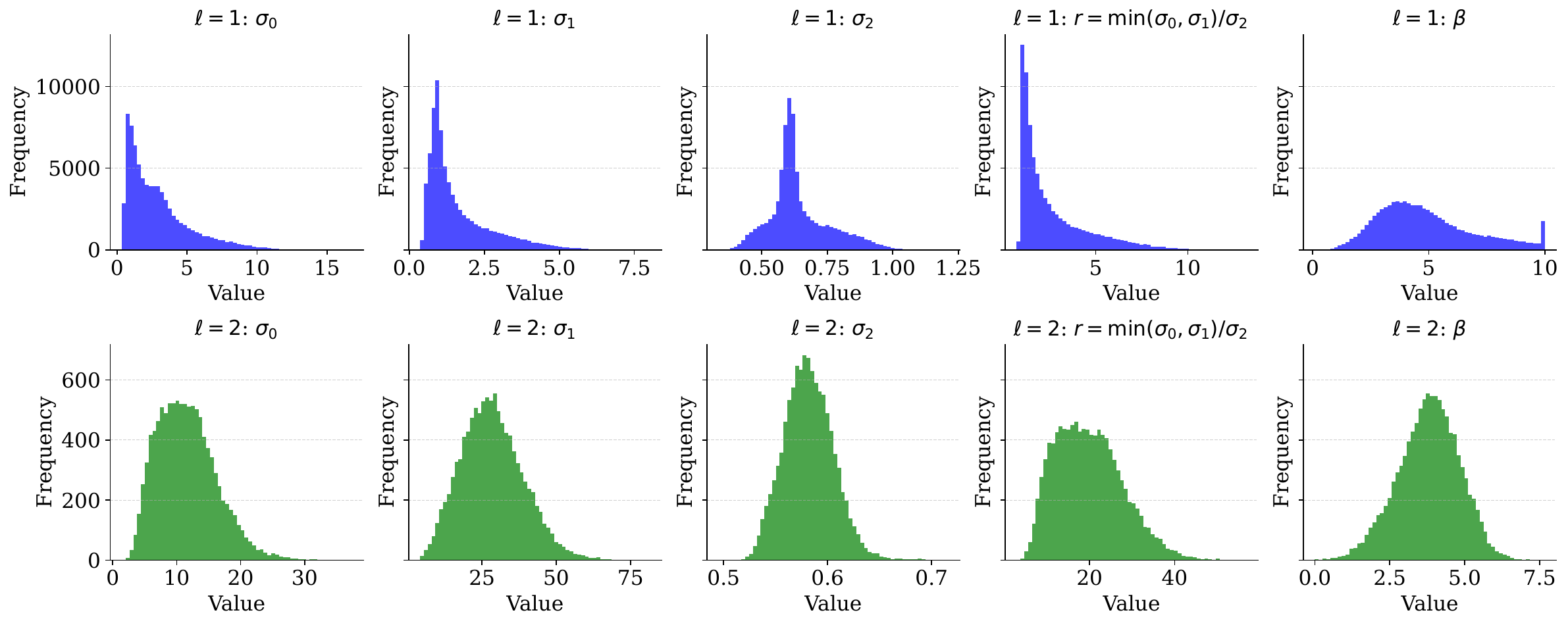}
  \caption{Distribution of pSurfel Parameters. $\sigma_i$: the $i$-th dimension of variance, $l$: layer index. $\beta$: shape coefficient.}
  \label{fig:vbeta}
\end{figure*}

\vspace{-2mm}\paragraph{Shape coefficient $\beta$.} We retrain by fixing $\beta=2$. We can witness a consistent performance drop when replacing the generalized Gaussian function by the Gaussian function.

\subsection{pSurfel Parameter Distribution}\label{abl:parameters}
% We visualize the histogram of the variance $\left\{\sigma_i\middle|i\in\left\{0,1,2\right\}\right\}$ and shape coefficient $\beta$ in Fig.~\ref{fig:vbeta}.

\vspace{-2mm}\paragraph{Variance:} At octree layer $l=2$, there are typically  two axes with large variances  ($\sigma_0$ and $\sigma_1$) and one with smaller variance ($\sigma_2$). For quantification, we define the planarity ratio
    $r=\min{(\sigma_0, \sigma_1)}/\sigma_2$,
where $\sigma_2$ corresponds to the smallest variance  (surfel thickness along normal). A high $r$ indicates that the pSurfel is effectively constrained to a thin surface. At $l=2$, the ratio $r$ is consistently large, confirming that $l=2$ pSurfels behave as planar structures bounded by the octree box. In contrast, at $l=1$, the variances across the three dimensions are more balanced, leading to smaller values of $r$. This suggests that the pSurfels at $l=1$ tend to approximate Gaussian-like blobs rather than thin planes, enabling the capture of non-planar local structures.

\vspace{-2mm}\paragraph{Shape coefficient $\beta$:}  $\beta$ exhibits a wider distribution at $l=1$. At this resolution, $\beta$ plays a stronger role in controlling the  drop-off sharpness of occupancy distribution, leading to greater variability. In contrast, at $l=2$, the pSurfels are coarser and often extend beyond the bounding box, making the influence of $\beta$  less significant.

\subsection{Visualization of Different Layers }
Fig.~\ref{fig:layers} visualizes the point cloud reconstructed from pSurfels at different layers of pSurfelTree. We pick a middle rate point where around half of the surfel nodes terminates at $l=2$ and the remaining at $l=1$. We can observe that pSurfels at coarser layers capture the smooth surfaces ({\it e.g.} the ground). pSurfels at finer layers usually represent high-frequency details with uneven surface normals ({\it e.g.} trees). This indicates that the decision module learns to distinguish between different geometry complexities and adaptively selects the  surfel granularity for rate-distortion trade-off.

\section{Conclusion}
This paper proposes the first end-to-end 
%, feed-forward 
surface-based point cloud geometry compression framework. It proposes a novel probabilistic surface representation termed pSurfel, which provides a differentiable likelihood over voxel occupancies. We further introduce a pSurfelTree structure  assigning  pSurfels  to different tree levels to reach optimal trade-off between rate and distortion. The whole network is end-to-end trained under the expected joint rate and distortion loss.
%We further introduce the hierarchical pSurfelTree structure for the adaptive tree decision under the rate-distortion trade-off. 
Experiments show significant improvement over voxel-based baselines.
%Our further work includes the introduction of color compression, and the color compression under the pSurfel representation.
Our future work will extend this work for color compression. In addition, we plan to explore SurfelSoup's extension to  LiDAR point clouds without explicit surface structures (See Appendix \ref{appendix:futurework}).

\bibliography{example_paper}

@article{wang2022sparse,
  title={Sparse tensor-based multiscale representation for point cloud geometry compression},
  author={Wang, Jianqiang and Ding, Dandan and Li, Zhu and Feng, Xiaoxing and Cao, Chuntong and Ma, Zhan},
  journal=PAMI,
  volume={45},
  number={7},
  pages={9055--9071},
  year={2022},
  publisher={IEEE}
}

@inproceedings{choy20194d,
  title={4D Spatio-Temporal ConvNets: Minkowski Convolutional Neural Networks},
  author={Choy, Christopher and Gwak, JunYoung and Savarese, Silvio},
  booktitle={CVPR},
  pages={3075--3084},
  year={2019}
}

@article{wang2024versatile1,
  title={A versatile point cloud compressor using universal multiscale conditional coding--part I: Geometry},
  author={Wang, Jianqiang and Xue, Ruixiang and Li, Jiaxin and Ding, Dandan and Lin, Yi and Ma, Zhan},
  journal={IEEE transactions on pattern analysis and machine intelligence},
  year={2024},
  publisher={IEEE}
}

@inproceedings{ijcai2022p126, title = {D-DPCC: Deep Dynamic Point Cloud Compression via 3D Motion Prediction}, author = {Fan, Tingyu and Gao, Linyao and Xu, Yiling and Li, Zhu and Wang, Dong}, booktitle = {Proceedings of the Thirty-First International Joint Conference on Artificial Intelligence, {IJCAI-22}}, publisher = {International Joint Conferences on Artificial Intelligence Organization}, editor = {Lud De Raedt}, pages = {898--904}, year = {2022}, month = {7}, note = {Main Track}, doi = {10.24963/ijcai.2022/126}, url = {https://doi.org/10.24963/ijcai.2022/126}, }

@inproceedings{wang2021multiscale,
  title={Multiscale point cloud geometry compression},
  author={Wang, Jianqiang and Ding, Dandan and Li, Zhu and Ma, Zhan},
  booktitle={2021 Data Compression Conference (DCC)},
  pages={73--82},
  year={2021},
  organization={IEEE}
}

@article{wang2023dynamic,
  title={Dynamic Point Cloud Geometry Compression Using Multiscale Inter Conditional Coding},
  author={Wang, Jianqiang and Ding, Dandan and Chen, Hao and Ma, Zhan},
  journal={arXiv preprint arXiv:2301.12165},
  year={2023}
}

@inproceedings{fu2022octattention,
  title={Octattention: Octree-based large-scale contexts model for point cloud compression},
  author={Fu, Chunyang and Li, Ge and Song, Rui and Gao, Wei and Liu, Shan},
  booktitle={Proceedings of the AAAI conference on artificial intelligence},
  volume={36},
  number={1},
  pages={625--633},
  year={2022}
}

@inproceedings{balle2017end,
  title={End-to-end optimized image compression},
  author={Ball{\'e}, Johannes and Laparra, Valero and Simoncelli, Eero P},
  booktitle={5th International Conference on Learning Representations, ICLR 2017},
  year={2017}
}

@inproceedings{balle2018variational,
  title={Variational image compression with a scale hyperprior},
  author={Ball{\'e}, Johannes and Minnen, David and Singh, Saurabh and Hwang, Sung Jin and Johnston, Nick},
  booktitle={International Conference on Learning Representations},
  year={2018}
}

@article{d20178i,
  title={8i voxelized full bodies-a voxelized point cloud dataset},
  author={d’Eon, Eugene and Harrison, Bob and Myers, Taos and Chou, Philip A},
  journal={ISO/IEC JTC1/SC29 Joint WG11/WG1 (MPEG/JPEG) input document WG11M40059/WG1M74006},
  volume={7},
  pages={8},
  year={2017}
}

@article{keming2018owlii,
  title={Owlii Dynamic human mesh sequence dataset},
  author={Keming, Cao and Yi, Xu and Yao, Lu and Ziyu, Wen},
  journal={Document ISO/IEC JTC1/SC29/WG11 m42816, San Diego},
  year={2018}
}

@article{kerbl20233d,
  title={3D Gaussian splatting for real-time radiance field rendering.},
  author={Kerbl, Bernhard and Kopanas, Georgios and Leimk{\"u}hler, Thomas and Drettakis, George},
  journal={ACM Trans. Graph.},
  volume={42},
  number={4},
  pages={139--1},
  year={2023}
}

@inproceedings{quach2020improved,
  title={Improved deep point cloud geometry compression},
  author={Quach, Maurice and Valenzise, Giuseppe and Dufaux, Frederic},
  booktitle={2020 IEEE 22nd International Workshop on Multimedia Signal Processing (MMSP)},
  pages={1--6},
  year={2020},
  organization={IEEE}
}

@article{wang2021lossy,
  title={Lossy point cloud geometry compression via end-to-end learning},
  author={Wang, Jianqiang and Zhu, Hao and Liu, Haojie and Ma, Zhan},
  journal={IEEE Transactions on Circuits and Systems for Video Technology},
  volume={31},
  number={12},
  pages={4909--4923},
  year={2021},
  publisher={IEEE}
}

@inproceedings{wang2025unipcgc,
  title={Unipcgc: Towards practical point cloud geometry compression via an efficient unified approach},
  author={Wang, Kangli and Gao, Wei},
  booktitle={Proceedings of the AAAI Conference on Artificial Intelligence},
  volume={39},
  number={12},
  pages={12721--12729},
  year={2025}
}

@article{xu2025point,
  title={Point Cloud Compression and Objective Quality Assessment: A Survey},
  author={Xu, Yiling and Zhang, Yujie and Xia, Shuting and Yang, Kaifa and Huang, He and Shan, Ziyu and Huang, Wenjie and Yang, Qi and Yang, Le},
  journal={arXiv preprint arXiv:2506.22902},
  year={2025}
}

@article{hu2025teso,
  title={TeSO: Representing and Compressing 3D Point Cloud Scenes with Textured Surfel Octree},
  author={Hu, Yueyu and Gong, Ran and Fan, Tingyu and Wang, Yao},
  journal={arXiv preprint arXiv:2508.07083},
  year={2025}
}

@article{maggiordomo2020real,
  title={Real-world textured things: A repository of textured models generated with modern photo-reconstruction tools},
  author={Maggiordomo, Andrea and Ponchio, Federico and Cignoni, Paolo and Tarini, Marco},
  journal={Computer Aided Geometric Design},
  volume={83},
  pages={101943},
  year={2020},
  publisher={Elsevier}
}

@INPROCEEDINGS{zhang2024standard,
  author={Zhang, Wei and Yang, Fuzheng and Xu, Yingzhan and Preda, Marius},
  booktitle={2024 Picture Coding Symposium (PCS)}, 
  title={Standardization Status of MPEG Geometry-Based Point Cloud Compression (G-PCC) Edition 2}, 
  year={2024},
  volume={},
  number={},
  pages={1-5},
  doi={10.1109/PCS60826.2024.10566443}}

@article{schwarz2018emerging,
  title={Emerging MPEG standards for point cloud compression},
  author={Schwarz, Sebastian and Preda, Marius and Baroncini, Vittorio and Budagavi, Madhukar and Cesar, Pablo and Chou, Philip A and Cohen, Robert A and Krivoku{\'c}a, Maja and Lasserre, S{\'e}bastien and Li, Zhu and others},
  journal={IEEE Journal on Emerging and Selected Topics in Circuits and Systems},
  volume={9},
  number={1},
  pages={133--148},
  year={2018},
  publisher={IEEE}
}

@inproceedings{que2021voxelcontext,
  title={Voxelcontext-net: An octree based framework for point cloud compression},
  author={Que, Zizheng and Lu, Guo and Xu, Dong},
  booktitle={Proceedings of the IEEE/CVF Conference on Computer Vision and Pattern Recognition},
  pages={6042--6051},
  year={2021}
}

@inproceedings{jang2017categorical,
  title = {Categorical Reparameterization with Gumbel-Softmax},
  author = {Eric Jang and Shixiang Gu and Ben Poole},
  booktitle = {International Conference on Learning Representations (ICLR)},
  year = {2017},
}

@article{minnen2018joint,
  title={Joint autoregressive and hierarchical priors for learned image compression},
  author={Minnen, David and Ball{\'e}, Johannes and Toderici, George D},
  journal={Advances in neural information processing systems},
  volume={31},
  year={2018}
}

@article{bengio2013estimating,
  title={Estimating or propagating gradients through stochastic neurons for conditional computation},
  author={Bengio, Yoshua and L{\'e}onard, Nicholas and Courville, Aaron},
  journal={arXiv preprint arXiv:1308.3432},
  year={2013}
}

@inproceedings{ICLR2025_804f1fac,
 author = {Huang, Tianxin and Yan, Zhiwen and Zhao, Yuyang and Lee, Gim H},
 booktitle = {International Conference on Representation Learning},
 editor = {Y. Yue and A. Garg and N. Peng and F. Sha and R. Yu},
 pages = {51765--51784},
 title = {ComPC: Completing a 3D Point Cloud with 2D Diffusion Priors},
 volume = {2025},
 year = {2025}
}

@inproceedings{zheng2024viewpcgc,
  title={Viewpcgc: view-guided learned point cloud geometry compression},
  author={Zheng, Huiming and Gao, Wei and Yu, Zhuozhen and Zhao, Tiesong and Li, Ge},
  booktitle={Proceedings of the 32nd ACM International Conference on Multimedia},
  pages={7152--7161},
  year={2024}
}

@inproceedings{ICLR2025_8e86fd92,
 author = {Jiang, Junzhe and Gu, Chun and Chen, Yurui and Zhang, Li},
 booktitle = {International Conference on Representation Learning},
 editor = {Y. Yue and A. Garg and N. Peng and F. Sha and R. Yu},
 pages = {56631--56645},
 title = {GS-LiDAR: Generating Realistic LiDAR Point Clouds with Panoramic Gaussian Splatting},
 volume = {2025},
 year = {2025}
}

@inproceedings{you2025reno,
  title={Reno: Real-time neural compression for 3d lidar point clouds},
  author={You, Kang and Chen, Tong and Ding, Dandan and Asif, M Salman and Ma, Zhan},
  booktitle={Proceedings of the Computer Vision and Pattern Recognition Conference},
  pages={22172--22181},
  year={2025}
}

@inproceedings{nguyen2021learning,
  title={Learning-based lossless compression of 3d point cloud geometry},
  author={Nguyen, Dat Thanh and Quach, Maurice and Valenzise, Giuseppe and Duhamel, Pierre},
  booktitle={ICASSP 2021-2021 IEEE International Conference on Acoustics, Speech and Signal Processing (ICASSP)},
  pages={4220--4224},
  year={2021},
  organization={IEEE}
}

@inproceedings{pfister2000surfels,
  title={Surfels: Surface elements as rendering primitives},
  author={Pfister, Hanspeter and Zwicker, Matthias and Van Baar, Jeroen and Gross, Markus},
  booktitle={Proceedings of the 27th annual conference on Computer graphics and interactive techniques},
  pages={335--342},
  year={2000}
}

@inproceedings{huang2020octsqueeze,
  title={Octsqueeze: Octree-structured entropy model for lidar compression},
  author={Huang, Lila and Wang, Shenlong and Wong, Kelvin and Liu, Jerry and Urtasun, Raquel},
  booktitle={Proceedings of the IEEE/CVF conference on computer vision and pattern recognition},
  pages={1313--1323},
  year={2020}
}

@article{biswas2020muscle,
  title={Muscle: Multi sweep compression of lidar using deep entropy models},
  author={Biswas, Sourav and Liu, Jerry and Wong, Kelvin and Wang, Shenlong and Urtasun, Raquel},
  journal={Advances in Neural Information Processing Systems},
  volume={33},
  pages={22170--22181},
  year={2020}
}

@article{zhou2018open3d,
  title={Open3D: A modern library for 3D data processing},
  author={Zhou, Qian-Yi and Park, Jaesik and Koltun, Vladlen},
  journal={arXiv preprint arXiv:1801.09847},
  year={2018}
}

@inproceedings{chao2025textured,
  title={Textured gaussians for enhanced 3d scene appearance modeling},
  author={Chao, Brian and Tseng, Hung-Yu and Porzi, Lorenzo and Gao, Chen and Li, Tuotuo and Li, Qinbo and Saraf, Ayush and Huang, Jia-Bin and Kopf, Johannes and Wetzstein, Gordon and others},
  booktitle={Proceedings of the Computer Vision and Pattern Recognition Conference},
  pages={8964--8974},
  year={2025}
}

@article{chang2015shapenet,
  title     = {ShapeNet: An Information-Rich 3D Model Repository},
  author    = {Chang, Angel X and Funkhouser, Thomas and Guibas, Leonidas and Hanrahan, Pat and Huang, Qixing and Li, Zimo and Savarese, Silvio and Savva, Manolis and Song, Shuran and Su, Hao and Xiao, Jianxiong and Yi, Li and Yu, Fisher},
  journal   = {arXiv preprint arXiv:1512.03012},
  year      = {2015}
}

@inproceedings{park2019deepsdf,
  title={Deepsdf: Learning continuous signed distance functions for shape representation},
  author={Park, Jeong Joon and Florence, Peter and Straub, Julian and Newcombe, Richard and Lovegrove, Steven},
  booktitle={Proceedings of the IEEE/CVF conference on computer vision and pattern recognition},
  pages={165--174},
  year={2019}
}

@inproceedings{fridovich2022plenoxels,
  title={Plenoxels: Radiance fields without neural networks},
  author={Fridovich-Keil, Sara and Yu, Alex and Tancik, Matthew and Chen, Qinhong and Recht, Benjamin and Kanazawa, Angjoo},
  booktitle={Proceedings of the IEEE/CVF conference on computer vision and pattern recognition},
  pages={5501--5510},
  year={2022}
}

@article{mildenhall2021nerf,
  title={Nerf: Representing scenes as neural radiance fields for view synthesis},
  author={Mildenhall, Ben and Srinivasan, Pratul P and Tancik, Matthew and Barron, Jonathan T and Ramamoorthi, Ravi and Ng, Ren},
  journal={Communications of the ACM},
  volume={65},
  number={1},
  pages={99--106},
  year={2021},
  publisher={ACM New York, NY, USA}
}

@article{chibane2020neural,
  title={Neural unsigned distance fields for implicit function learning},
  author={Chibane, Julian and Pons-Moll, Gerard and others},
  journal={Advances in Neural Information Processing Systems},
  volume={33},
  pages={21638--21652},
  year={2020}
}

@inproceedings{dai2024high,
  title={High-quality surface reconstruction using gaussian surfels},
  author={Dai, Pinxuan and Xu, Jiamin and Xie, Wenxiang and Liu, Xinguo and Wang, Huamin and Xu, Weiwei},
  booktitle={ACM SIGGRAPH 2024 conference papers},
  pages={1--11},
  year={2024}
}

@inproceedings{gao2023surfelnerf,
  title={Surfelnerf: Neural surfel radiance fields for online photorealistic reconstruction of indoor scenes},
  author={Gao, Yiming and Cao, Yan-Pei and Shan, Ying},
  booktitle={Proceedings of the IEEE/CVF Conference on Computer Vision and Pattern Recognition},
  pages={108--118},
  year={2023}
}

@inproceedings{aliev2020neural,
  title={Neural point-based graphics},
  author={Aliev, Kara-Ali and Sevastopolsky, Artem and Kolos, Maria and Ulyanov, Dmitry and Lempitsky, Victor},
  booktitle={ECCV},
  pages={696--712},
  year={2020},
  organization={Springer}
}

@inproceedings{zwicker2001surface,
  title={Surface splatting},
  author={Zwicker, Matthias and Pfister, Hanspeter and Van Baar, Jeroen and Gross, Markus},
  booktitle={Proceedings of the 28th annual conference on Computer graphics and interactive techniques},
  pages={371--378},
  year={2001}
}

@article{kobbelt2004survey,
  title={A survey of point-based techniques in computer graphics},
  author={Kobbelt, Leif and Botsch, Mario},
  journal={Computers \& Graphics},
  volume={28},
  number={6},
  pages={801--814},
  year={2004},
  publisher={Elsevier}
}

@inproceedings{botsch2005high,
  title={High-quality surface splatting on today's GPUs},
  author={Botsch, Mario and Hornung, Alexander and Zwicker, Matthias and Kobbelt, Leif},
  booktitle={Proceedings Eurographics/IEEE VGTC Symposium Point-Based Graphics, 2005.},
  pages={17--141},
  year={2005},
  organization={IEEE}
}

@inproceedings{luo2024scp,
  title={Scp: Spherical-coordinate-based learned point cloud compression},
  author={Luo, Ao and Song, Linxin and Nonaka, Keisuke and Unno, Kyohei and Sun, Heming and Goto, Masayuki and Katto, Jiro},
  booktitle={Proceedings of the AAAI Conference on Artificial Intelligence},
  volume={38},
  number={4},
  pages={3954--3962},
  year={2024}
}

@inproceedings{han2016deep,
  title={Deep Compression: Compressing Deep Neural Network with Pruning, Trained Quantization and Huffman Coding},
  author={Han, Song and Mao, Huizi and Dally, William J},
  booktitle={ICLR},
  year={2016}
}

@inproceedings{chen2025knowledge,
  title={Knowledge Distillation for Learned Image Compression},
  author={Chen, Yunuo and Lyu, Zezheng and He, Bing and Cao, Ning and Chen, Gang and Lu, Guo and Zhang, Wenjun},
  booktitle={Proceedings of the IEEE/CVF International Conference on Computer Vision},
  pages={4996--5006},
  year={2025}
}

@inproceedings{dai2017scannet,
  title={Scannet: Richly-annotated 3d reconstructions of indoor scenes},
  author={Dai, Angela and Chang, Angel X and Savva, Manolis and Halber, Maciej and Funkhouser, Thomas and Nie{\ss}ner, Matthias},
  booktitle={Proceedings of the IEEE conference on computer vision and pattern recognition},
  pages={5828--5839},
  year={2017}
}
\bibliographystyle{icml2026}

%%%%%%%%%%%%%%%%%%%%%%%%%%%%%%%%%%%%%%%%%%%%%%%%%%%%%%%%%%%%%%%%%%%%%%%%%%%%%%%
%%%%%%%%%%%%%%%%%%%%%%%%%%%%%%%%%%%%%%%%%%%%%%%%%%%%%%%%%%%%%%%%%%%%%%%%%%%%%%%
% APPENDIX
%%%%%%%%%%%%%%%%%%%%%%%%%%%%%%%%%%%%%%%%%%%%%%%%%%%%%%%%%%%%%%%%%%%%%%%%%%%%%%%
%%%%%%%%%%%%%%%%%%%%%%%%%%%%%%%%%%%%%%%%%%%%%%%%%%%%%%%%%%%%%%%%%%%%%%%%%%%%%%%
\newpage
\appendix
\onecolumn
\section{Appendix for Methods}
\begin{table}[t]
\centering
\footnotesize
\caption{Summary of Notation.}
\label{tab:notation}
\begin{tabular}{ll}
\toprule
\textbf{Symbol} & \textbf{Description} \\
\midrule
$L$ & Maximum number of downsampling in feature encoding (set as 3) \\
$l\in\left\{0,\cdots,L\right\}$ & Octree level index, $0$ denotes the finest level, $L$ the coarsest level. \\
%$i, j$ & Node and octant indices \\
${\bf u}_i^l$ & A node $i$ at level $l$ \\
${\bf u}_{i,j}^l$ & The $j$-th octant of node $\mathbf{u}_i^l$ \\
$\hat{\mathbf{f}}_i^l$ & Latent feature of ${\bf u}_i^l$ \\
$\hat{\bf f}_{i,j}^l$ & Latent feature of octant ${\bf u}_{i,j}^l$ \\
$\mathbf{s}_i^l$ & ${\bf u}_i^l$ classfied as a surfel node \\
$\mathbf{o}_i^l$ & ${\bf u}_i^l$ classfied as an octree node \\
${\bf v}_{i,k}^l$ & the $k$-th voxel (point with integer coordinate) in $\mathbf{s}_i^l$ \\
$\mathcal{V}_i^l$ & The set of voxels ${\bf v}_{i,k}^l$ in the bounding box defined by $\mathbf{s}_i^l$ \\
$\mu,\sigma,q,\beta$ & Mean, variance, quaternion and shape of pSurfels \\
$\mathcal{D}(\mathbf{s}_i^l)$ & Distortion incurred by representing ${\bf u}_i^l$ as a surfel  \\
$\pi(i,l)$ & The index of the parent of ${\bf u}_i^l$ in layer $l+1$\\
$\tilde{p}_i^l$ &  Probability of ${\bf u}_i^l$ being an octree node estimated by the decision module \\
$\tilde{q}_i^l$ & Probability of ${\bf u}_i^l$ being a surfel node estimated by the decision module \\
$\mathcal{O}_{i,j}^l \in \{0,1\}$ & Ground truth occupancy of $j$-th octant under node $i$, $\mathbf{u}_{i,j}^l$ \\
$\mathcal{O}_i^l \in \{0,1\}^8$ & Occupancy of 8 octants under node ${\bf u}_i^l$, $\mathcal{O}_i^l=\left\{ \mathcal{O}_{i,j}^l|j\in\left[:8\right]\right\}$ \\
$\mathcal{O}_{\left[j\right]}^l$ & The set of octants in layer $l$ with group index $j$, $\mathcal{O}_{\left[j\right]}^l=\left\{\mathcal{O}_{i,j}^l|\forall i\right\}$ \\
$\mathcal{O}_{\left[:j-1\right]}^l$ & The set of octants in layer $l$ with group index $j'<j$, $\mathcal{O}_{\left[:j-1\right]}^l = \left\{\mathcal{O}_{i,j}|\forall i;j'\in\left[:j-1\right]\right\}$\\
$\theta_{i,j}^l$ & P-SOPA-predicted probability that $\mathcal{O}_{i,j}^l=1$\\
$\theta_{\left[j\right]}^l$ & P-SOPA-predicted probabilities of group $j$ octants, $\theta_{\left[j\right]}^l=\left\{\theta_{i,j}^l|\forall i\right\}$\\ %\mathcal{O}_{\left[j\right]}^l$\\
$\mathcal{P}_{\left[:j-1\right]}^l$ & Probability of existence of $\mathcal{O}_{\left[:j-1\right]}^l$, $\mathcal{P}_{\left[:j-1\right]}^l=\left\{\mathcal{P}_{i,j'}^l=\tilde{p}_i^l|\forall i;j'\in\left[:j-1\right]\right\}$ \\
$m_{\left[:j-1\right]}^l$ & Bernoulli sampled mask based on $\mathcal{P}_{\left[:j-1\right]}^l$ \\
$\bar{m}_{\left[:j-1\right]}^l$ & STE version of  $m_{\left[:j-1\right]}^l$ \\

$\mathcal{B}(\mathbf{o}_i^l)$ & Bit rate estimation of coding the occupancy of octants of $\mathbf{o}_i^l$, $\mathcal{O}_{i}^l$ \\
$\epsilon$ & Threshold for surfel termination during evaluation \\
$\mathrm{Bernoulli}(\cdot)$ & Bernoulli distribution \\
\bottomrule
\vspace{-2mm}
\end{tabular}
\end{table}

 \vspace{-2mm}\begin{figure}[tbp]
  \centering
    \includegraphics[width=0.9\linewidth]{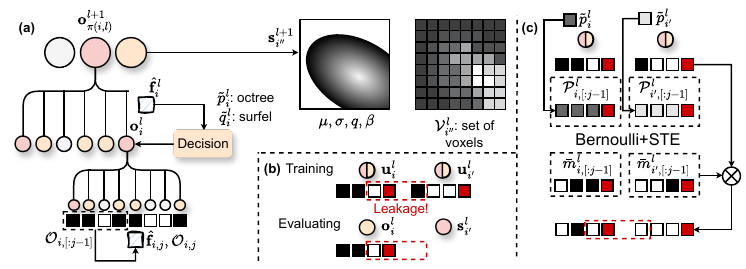}
   \caption{(a) Illustrations of notations. (b) a toy example of information leakage during training. The red dashed box indicates convolution, only four octants out of the eight octants are drawn. Black/white/red indicate that an octant is occupied/unoccupied/unknown (not decoded yet).  The decoding of groups is conducted from left to right. When decoding the last group of ${\bf u}_i^l$, we apply a convolution (red dashed line) on the nearby already decoded groups for context generation. However, during evaluation, some decoded octants may not exist ({\it e.g.} ${\bf u}_{i'}^l$'s first node) because its parent is classified as a surfel (${\bf s}_{i'}^l$), while during training the network can still access it because no actual termination is performed.(c) a toy example with only two nodes of P-SOPA. The color of $\tilde{p}^l$ and $\mathcal{P}_{\left[:j-1\right]}^l$ indicates the value of probability, with darker color indicating higher probability of existence. We sample a Bernoulli-STE mask $\bar{m}^l_{\left[:j-1\right]}$ from $\mathcal{P}^l_{\left[:j-1\right]}$ to prune the already decoded octants, so that the training better simulates the actual decoding.}
   \label{fig:notations}
\end{figure}

\subsection{Notations}
\label{appendix:notation}
We summarize the main notation used throughout the paper in Tab.~\ref{tab:notation}. For better understanding, we further provide an illustration of each notation in Fig.~\ref{fig:notations} (a).

\subsection{More Information on P-SOPA}\label{app:psopa}
We give a toy example about why we  care about the information leakage during training in the middle of Fig.~\ref{fig:notations}, with explanations why information leakage occur in the caption. Therefore, to make training more close to real decoding, we apply a Bernoulli-STE mask $\bar{m}^l_{\left[:j-1\right]}$ on the already decoded octants $\mathcal{O}^l_{\left[:j-1\right]}$, to simulate the process during evaluation that some octants in $\mathcal{O}^l_{\left[:j-1\right]}$ do not exist, because their parents are classified as surfel nodes (which are no further divided). This mask, $\bar{m}^l_{\left[:j-1\right]}$, is generated by duplicating the parents' probability of being classified as octree nodes, {\it i.e.} $\tilde{p}^l$. Because octants exist iff. their parents are octree nodes that need further division.

\subsection{Network Structures}\label{appendix:ns}
This section introduces the network architecture of SurfelSoup, including the encoder module (Sec. 3.2), pSurfel Reconstruction (Sec. 3.3), decision module (Sec. 3.4) and P-SOPA (Sec. 3.5).
\subsubsection{Encoder}
The encoder architecture is shown in Fig.~\ref{fig:detailed} (a). We follow the encoder design in SparsePCGC~\citep{wang2022sparse} and Unicorn~\citep{wang2024versatile1}, consisting of two ResNet-3 Modules~\citep{wang2024versatile1} and one stride-two sparse convolution layer for downsampling.

\subsubsection{P-SOPA}
The detailed architecture of P-SOPA is shown in Fig.~\ref{fig:detailed} (c). P-SOPA follows the design of the SOPA module in SparsePCGC~\citep{wang2022sparse}. However, the ResNet-3 is replaced by P-ResNet-3, where an STE mask in Sec.~\ref{sec:psopa} is applied to the input of every convolution layer.

\subsubsection{Decision/pSurfel Reconstructor}
\begin{figure}[htbp]
  \centering
    \includegraphics[width=0.7\linewidth]{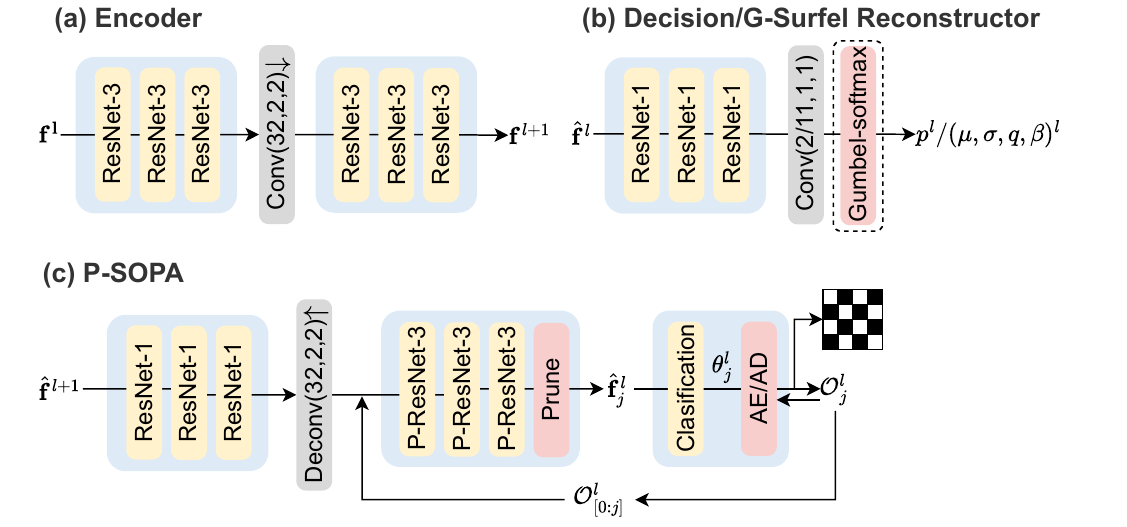}
   \caption{The architecture for (a) Encoder, (b) Decision Module/pSurfel Reconstructor and (c) P-SOPA. ResNet-$x$ denotes the Sparse ResNet structure~\citep{wang2024versatile1} constructed by stride-$x$ convolutions. Conv($C$,$K$,$S$) denotes convolution with channel size $C$, kernel size $K$ and stride $S$. The dashed-line box indicates that Gumbel-Softmax is only for decision module. {\it AE/AD} denotes the arithmetic encoder and arithmetic decoder.} 
   \label{fig:detailed}
\end{figure}

The decision module/pSurfel Reconstructor is shown in Fig.~\ref{fig:detailed} (b), with one ResNet-1 Module and a stride-one sparse convolution module that changes the output to the desired dimension. For decision module, there is an additional Gumbel-Softmax. Note that we use stride-one instead of stride-three convolutions to avoid the information leakage mentioned in Sec. 3.5.

\begin{figure}[tbp]
  \centering
    \includegraphics[width=0.8\linewidth]{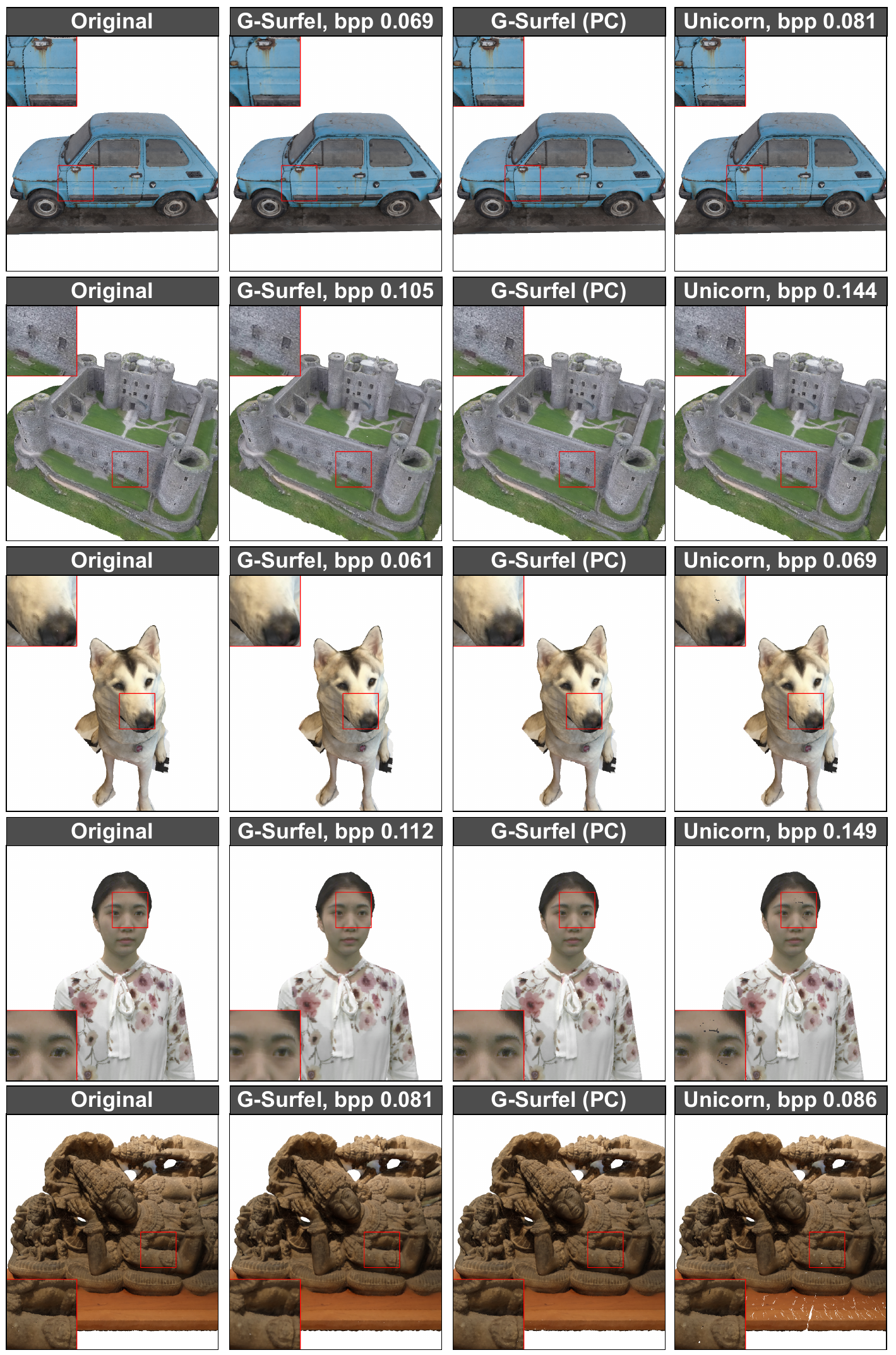}
   \caption{Comparison between rendered images from the decoded point clouds using different methods.  Column pSurfel denotes  visualization of decoded surfels by SurfelSoup.  Column pSurfel (PC) denotes  visualization of decoded points by SurfelSoup using OpenGL. Unicorn indicates the rendered images of decoded points by Unicorn using OpenGL. The color of a decoded point is obtained  by interpolating on the colors of the original point cloud. Row 1, 2 and 3 are three samples from RWTT~\citep{maggiordomo2020real}. } 
   \vspace{-2mm}
   \label{fig:morevis}
\end{figure}

\paragraph{pSurfel parameters.} A pSurfel is defined by its mean $\mu\in\mathbb{R}^3$, variance $\sigma\in\mathbb{R}^3$, quaternion $q\in\mathbb{R}^4$ and shape coefficient $\beta\in\mathbb{R}^3$. The rotation matrix is derived by $q=\left[w,x,y,z\right]$:
\begin{equation}
R(q)=
\begin{bmatrix}
w^2+x^2-y^2-z^2 & 2(xy-wz)         & 2(xz+wy) \\
2(xy+wz)         & w^2-x^2+y^2-z^2 & 2(yz-wx) \\
2(xz-wy)         & 2(yz+wx)        & w^2-x^2-y^2+z^2
\end{bmatrix}.
\end{equation}
The covariance matrix is derived by the rotation matrix $R(q)$ and variance $\sigma$:
\begin{equation}
\Sigma \;=\; R\,\mathrm{diag}(\sigma_0^2,\,\sigma_1^2,\,\sigma_2^2)\,R^\top,
\qquad
\boldsymbol{\varepsilon} \;=\; \Sigma^{-1}
\;=\; R\,\mathrm{diag}\!\big(\sigma_0^{-2},\,\sigma_1^{-2},\,\sigma_2^{-2}\big)\,R^\top.
\end{equation}

\subsection{Comparison with Other 3D Representation}\label{appendix:compareGS}
It is worth clarifying how SurfelSoup differs from recent 3D representation techniques such as Gaussian splatting, neural implicit fields, signed distance
functions, and surfels. These methods aim to build continuous or differentiable scene representations from multi-view supervision and per-scene optimization. These methods often targets at the
quality of novel-view rendering, and often have huge amount of parameters ($>$20MB) because no
rate constraint is applied. In contrast, our setting focuses on point cloud geometry compression,
where the goal is to encode the input 3D point cloud into a compact bitstream ($<$0.05MB) under
standardized rate–distortion constraints, while ensuring the quality of the 3D reconstructed point
cloud. For concreteness, we provide a qualitative comparison with 3D Gaussian Splatting in Tab.~\ref{tab:compare_3dgs}, while the same mismatch in assumptions and objectives applies broadly to other 3D representation families.

\begin{table}[h]
\centering
\small     % ← 调整字体大小：\footnotesize 或 \scriptsize 也可以
\caption{Comparison between SurfelSoup and 3D Gaussian Splatting (3DGS).}
\label{tab:compare_3dgs}
\begin{tabular}{lcc}
\toprule
\textbf{Property} & \textbf{SurfelSoup} & \textbf{3DGS} \\
\midrule
Representation primitives & Probabilistic pSurfels & 3D Gaussians \\
Structure & Hierarchical pSurfelTree & unstructured set of Gaussians \\
Training paradigm & Dataset-level learning & Per-scene optimization \\
Objective & Rate--distortion optimization & Photometric rendering loss \\
Representation target & Geometry only & Geometry + color + opacity \\
Compressibility & Built-in entropy coding & External compression required \\
Granularity control & Learned adaptive decision & Hand-crafted prune/densify \\
Scene generalization & Yes  & Scene-specific optimization  \\
Primary use-case & Point cloud geometry compression & 3D reconstruction for rendering \\
Speed & Feed-forward inference & Per-scene optimization \\
\bottomrule
\end{tabular}
\end{table}

\section{Appendix for Experiments}
\subsection{Surfel Rendering.}\label{appendix:rendering}

In Fig.~\ref{fig:morevis}, we compare rendered images by three methods: OpenGL point cloud rendering for SurfelSoup reconstructed points, Our bounded 2D surfel based rendering for surfelSoup directly, and OpenGL point cloud rendering for  Unicorn.

\paragraph{OpenGL Rendering.}
In OpenGL~\citep{zhou2018open3d} rendering, each point is splatted on to the screen using a fixed $n \times n$ pixel patch, where $n$ is manually set. To ensure fairness, we determine $n$ by selecting the minimum number that eliminates visible gaps in the rendered images from the ground-truth uncompressed point cloud. This chosen point size is then applied to both pSurfel (PC) and Unicorn. 
%fixed across all point cloud compression methods to enable consistent visual comparison.

\paragraph{Surfel Rendering}
In addition to rendering the decoded point clouds by SurfelSoup using OpenGL, we can also render the decoded surfels  directly. For each cube decoded to a pSurfel, we identify the direction with the minimal variance and use it as the normal direction of the surfel plane.  
%we assign the normal and scale of the surfel to the cube to define a local tangent plane. The surfel normal is chosen along its smallest scaling dimension,  and the other two tangent plane axes are obtained by a rotation aligning the global $z$-axis with the normal. 
We define the  two axes of the surfel by   a rotation aligning the global $z$-axis with the normal. 
%We assign a texture map to each pSurfel by treating it as a 2D surfel. The reconstructed point cloud is spatially organized into an octree. 
We construct an $n \times n$ texture map on this tangent plane, where $n$ varies with the octree level, so that it covers the largest possible surfel bounded by the cube. All decoded points within the cube and its neighbors are projected to this plane, which ensures consistency of texture samples across neighboring cubes. For each pixel in the texture map, we query the k-nearest projected points and assign the pixel a color equal to the inverse-distance-weighted average of these k neighbors.  We use $k=3$, $n=7$ for $l=1$, $n=15$ for $l=2$.

For rendering, we perform rasterization from the given camera view. We first sort all potential surfels for a pixel by their distances to the camera center. Then, along each camera ray, we compute the intersection point between the ray and the surfel surface. If this point lies inside the surfel's bounding box, we convert this point into the tangent plane coordinate and query the corresponding texture map using bilinear interpolation to shade the pixel from the texture patch. Otherwise, we proceed to the next surfel until a valid hit is found. This design enables efficient CUDA-based rasterization, follows the logic of textured 2D Gaussians\citep{chao2025textured}, but leverages a bounded surfel representations\citep{hu2025teso}. 

\begin{figure}[tbp]
  \centering
    \includegraphics[width=0.55\linewidth]{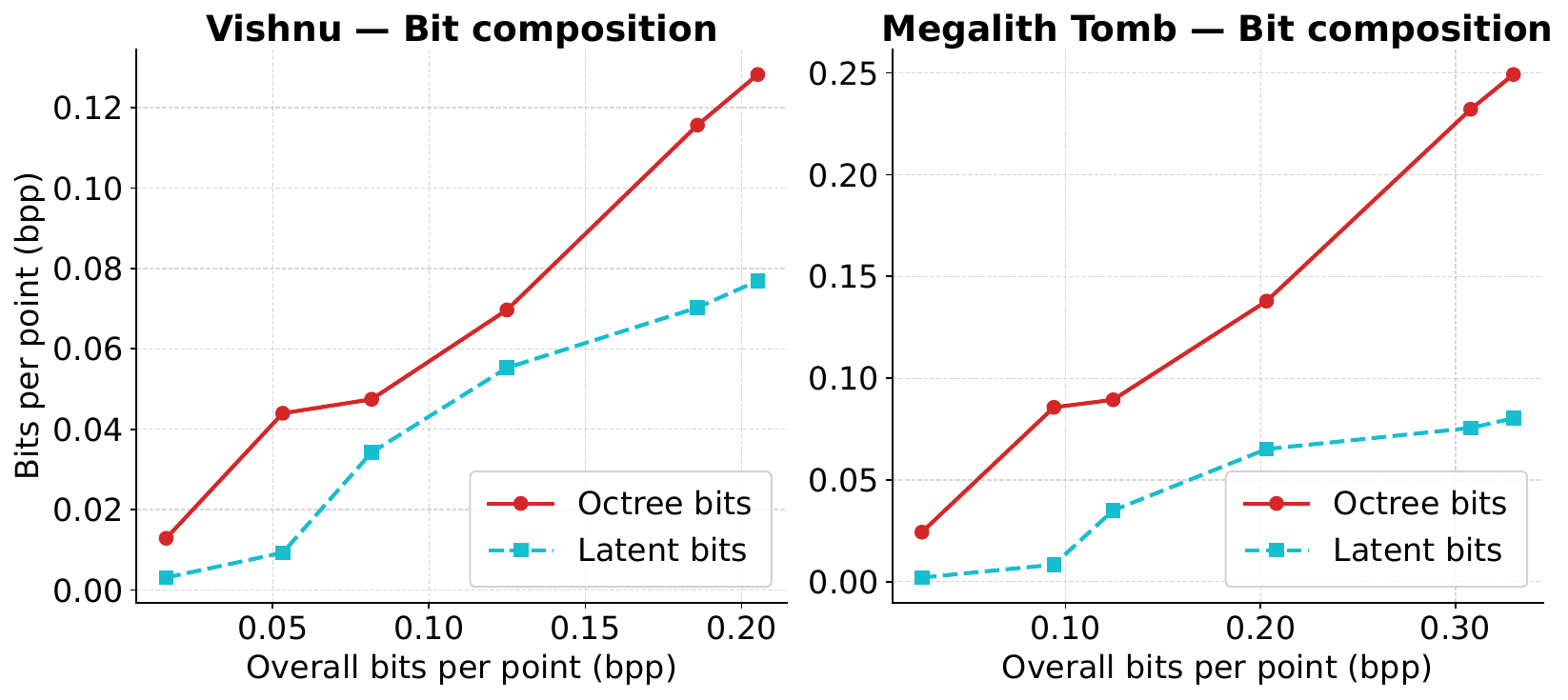}
    \vspace{-2mm}
   \caption{Bit rate composition (octree bits and latent bits).} 
   \vspace{-3mm}
   \label{fig:composition}
\end{figure}

\paragraph{Rendering Result}
We can observe that at similar geometry rates, both pSurfel and pSurfel (PC) significantly outperform Unicorn, which suffer from visible gaps in some areas. Compared with the pSurfel renders using OpenGL, surfel rendering provides slightly more smooth color and geometry distribution (the readers may need to further zoom in).

\begin{figure}[tbp]
  \centering
    \includegraphics[width=0.95\linewidth]{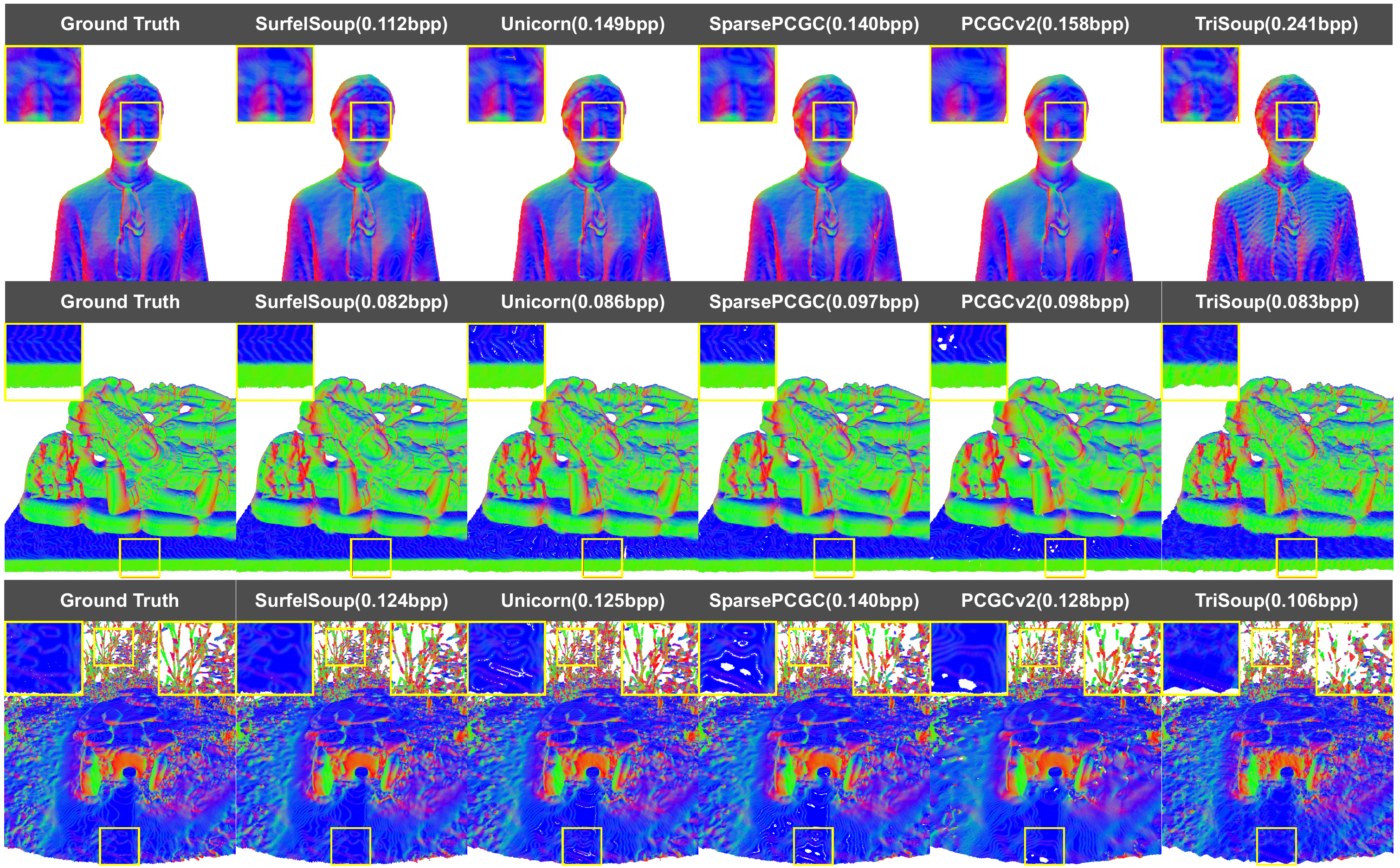}
    \vspace{-2mm}
   \caption{Visual comparison between SurfelSoup and baselines. The point color is visualized by the normal vector.} 
   \vspace{-3mm}
   \label{fig:comparegeo}
\end{figure}

\subsection{Test Results on ScanNet}\label{appendix:scannet}
\begin{figure}[tbp]
  \centering
    \includegraphics[width=0.95\linewidth]{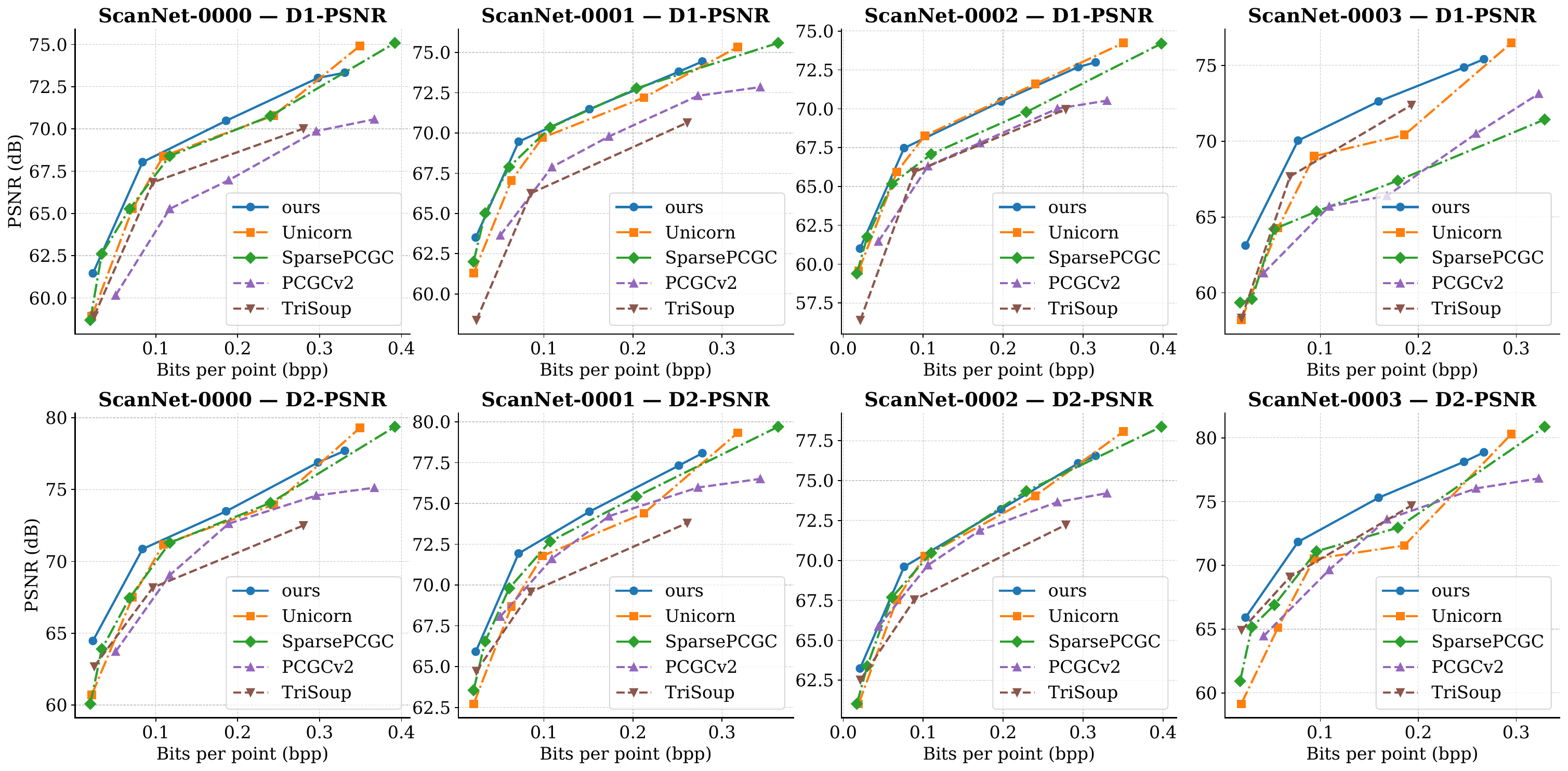}
    \vspace{-2mm}
   \caption{RD-curves on ScanNet-0001$\rightarrow$ -0003.} 
   \vspace{-3mm}
   \label{fig:rdscannet}
\end{figure}

To validate our model's generalization ability to scene point clouds, we further provide comparison on ScanNet dataset {\bf without finetuning}. We test on the first four sequences, and the result is shown in Fig.~\ref{fig:rdscannet}. Our model, trained on human data, shows great generalization on scene point clouds, and still outperforms other baselines.

\subsection{Bits Composition}
We visualize the bit rate composition (bpp for octree compression and bpp for latent compression) in Fig.~\ref{fig:composition}. Compared with voxel baselines~\citep{wang2021multiscale,wang2024versatile1,wang2022sparse} which codes the entire octree  to some level, our method supports terminating a portion of the octree as surfels through the decision module. This gives the network great flexibility in adaptively selecting data structures (voxels or surfels) for geometric representation. 

We can further observe that: 1) the octree takes more bits as total rate goes higher, indicating that more nodes in coarser layers are represented as octree nodes for better quality surfels in higher layers; and 2) compared with Megalith Tomb, the percentage of octree bits of Vishnu is lower, indicating that Vishnu has more smooth surface structures. This aligns with the fact that Megalith Tomb is an irregular scene point cloud, while Vishnu is a dense object point cloud.

\subsection{Visual Comparison}\label{appendix:geoonly}
We compare the surface normal visualization between SurfelSoup and baselines in Fig.~\ref{fig:comparegeo}. Different from Fig.~\ref{fig:visual}, the visualization in this section does not carry color information. Because human eyes are more sensitive to distortion when color is rendered, we put the pure geometry visualization in the Appendix.

\subsection{Additional Results}\label{sec:thai}
\begin{figure}[htbp]
  \centering
    \includegraphics[width=0.55\linewidth]{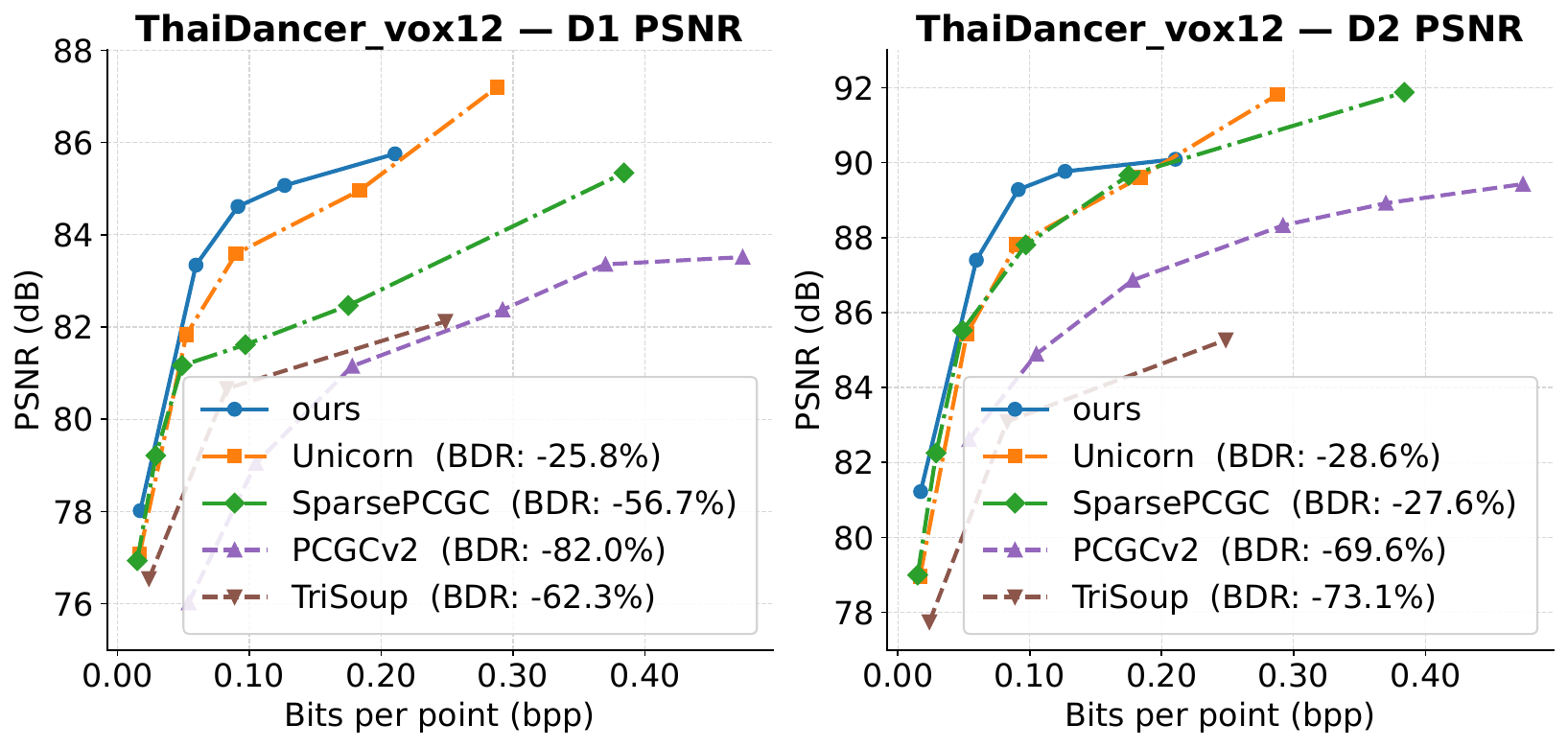}
    \vspace{-2mm}
   \caption{Comparison of SurfelSoup and baselines on {\it ThaiDancer\_vox12}.} 
   \vspace{-3mm}
   \label{fig:thai}
\end{figure}

\paragraph{ThaiDancer:} To further demonstrate the generalizability of our models trained on the 10-bit 8iVFB dataset, Fig.~\ref{fig:thai} shows the testing result on the 12-bit MPEG CTC test sequence {\it ThaiDancer\_vox12}. 
% It outperformed Unicorn signficantly over an intermediate rate range, but is inferior at the higher rates.

\begin{figure}[tbp]
  \centering
    \includegraphics[width=0.9\linewidth]{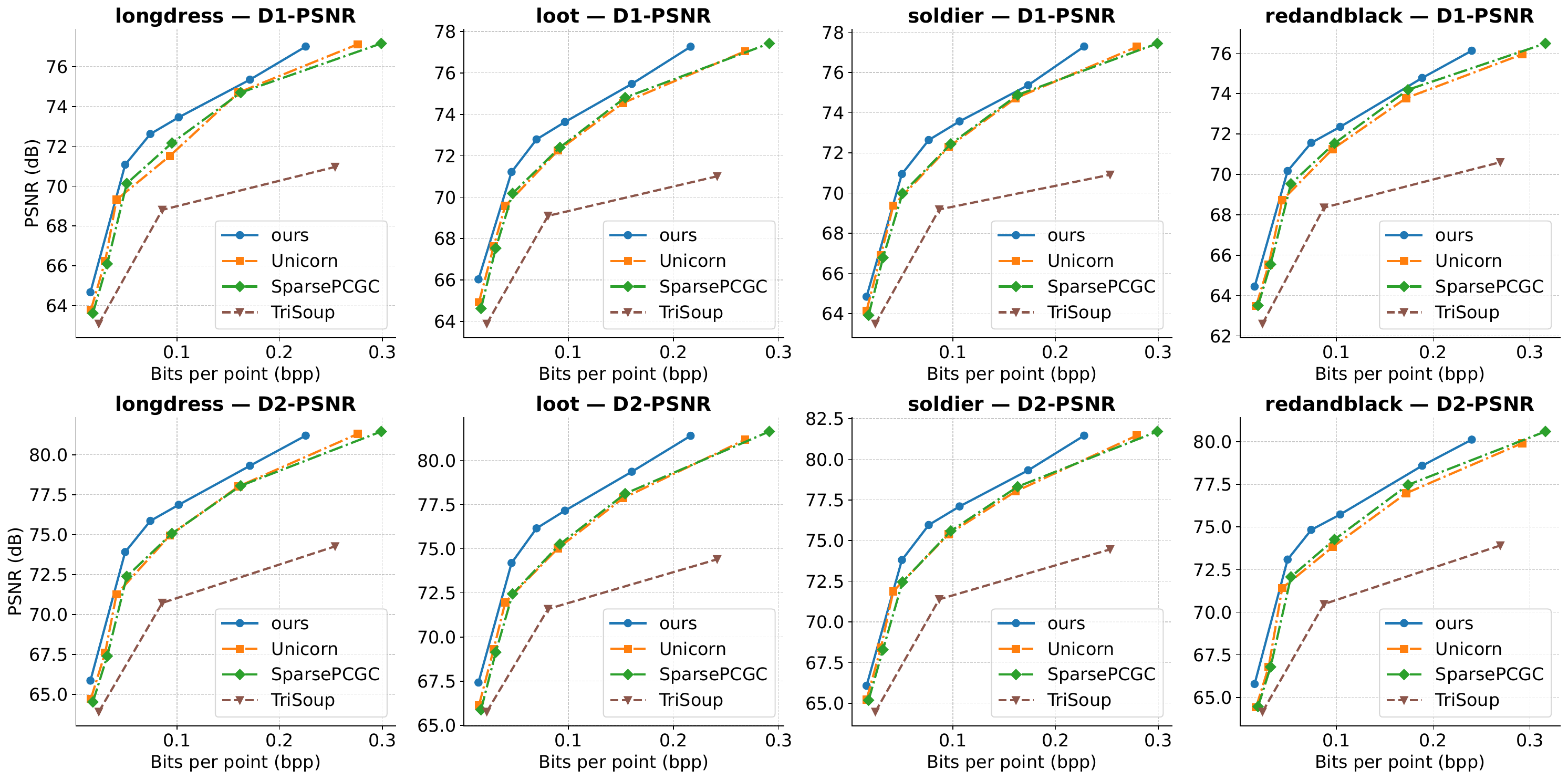}
    \vspace{-2mm}
   \caption{Comparison of SurfelSoup and baselines on 8iVFB. SurfelSoup and SparsePCGC are retrained on RWTT.} 
   \vspace{-3mm}
   \label{fig:8i}
\end{figure}

\paragraph{8iVFB:} To provide test result on 8iVFB, we retrain our model on another MPEG CTC-specified training set: RWTT~\citep{maggiordomo2020real}. The result is provided in Fig.~\ref{fig:8i}. We compare with Unicorn and SparsePCGC by retraining SparsePCGC also on RWTT. We do not compare with PCGCv2 because its performance is shown to be worse than Unicorn and SparsePCGC in previous studies. In addition, we compare with MPEG standard G-PCC-TriSoup. SurfelSoup consistently outperformed both SparsePCGC and G-PCC-TriSoup over the entire rate range.

\subsection{Training Details}
We train five models with $\lambda=0.1,0.3,0.8,1.0,1.5$. The initial learning rate is set as $0.0001$, which is multiplied by 0.85 after every 15 epochs. Batch size is set as 1. The model for each rate point is trained for one day. The lowest $\lambda$ ($\lambda=0.1$) corresponds to $\text{bpp}\approx0.05$. To obtain even lower rate points, we adopt the super-resolution approach in SparsePCGC~\citep{wang2022sparse} and Unicorn~\citep{wang2024versatile1}. Specifically we downsample the original point cloud by a stride-2 pooling layer, and compress this down-sampled point cloud by our trained model with $\lambda=0.1$. We then  leverage a super-resolution network \cite{wang2024versatile1} to upsample the decoded point cloud  to the original scale.

\paragraph{Model Configuration.} The number of down-sampling stages $L$ is set to 3. The geometry coordinates up to layer $L=3$ are losslessly coded with G-PCC-Octree. 
%The overall number of layers $L$ is set as 3, because the bit rate of losslessly coding layer $l>3$'s geometry with G-PCC-Octree is negligible. 
% Since encoding the octants' occupancy information of nodes at $l=3$ requires an extremely low bitrate ($<0.015$ bpp), performing pSurfel reconstruction at this level becomes suboptimal under the rate–distortion trade-off. Therefore, nodes at $l=3$ are forced to be classified as octree nodes.
Furthermore, given the excessive complexity and bit rate to code the occupancy information at $l=0$,  nodes at $l=1$ are forced to be classified as surfel nodes.

Due to the different convergence speeds of different octree layer's pSurfel Reconstruction (finer layer converges faster because it has less points inside each surfel node), the network cannot be trained from scratch. Otherwise, the  decision module converges to a local minimum, which classifies all  the nodes in coarser layers  as octree nodes. Instead, we first pre-train a  model by setting $\tilde{q}_i^l=0.5, \forall i,$ and $\lambda=2.0$ to better support the convergence of different layers' pSurfel Reconstruction. We then fine-tune this pre-trained model with different $\lambda.$

For Gumbel Softmax, the temperature coefficient is set as $\tau=1.5\times \exp(-0.0002\times\text{step})$.

\subsection{Complexity}

\paragraph{Encoding/Decoding Time.}  As shown in Table~\ref{tab:complexity_comparison_transposed}, despite the introduction of pSurfel, our encoding/decoding time is almost identical compared with Unicorn and SparsePCGC, because the bulk of time is on SOPA's auto-regressive coding of octree nodes.

\paragraph{Model Size.} Our model has larger size compared with Unicorn due to  the introduction of pSurfel Reconstruction and decision module (Table~\ref{tab:complexity_comparison_transposed}). 
% 2) Unicorn shares SOPA's model weights across layers. We have confirmed with Unicorn’s authors that Unicorn does not benefit greatly from non-shareable layers. Conversely, using independent weights for each layer is essential for our method, because results (Fig~\ref{fig:layers}, \ref{fig:vbeta}) show that different layers' pSurfels have very different distributions.
However, the overall model size is only moderately larger than that of Unicorn. In addition, the model size can be further reduced before real applications, with the pruning, distillation and quantization of model parameters (See Appendix~\ref{appendix:futurework}).

To better contextualize the model size difference, we additionally evaluate Unicorn with an increased channel width, enlarging the channel number from 32 to 48. This modification increases Unicorn’s model size from 22 MB to 49 MB, making it comparable to or even larger than SurfelSoup. However, this larger Unicorn variant does not lead to improved rate–distortion performance. Instead, we observe a {\bf 0.6\% BD-rate degradation} compared to the original 32-channel Unicorn. This result indicates that simply scaling up the network capacity does not necessarily improve compression efficiency, and suggests that the performance gain of SurfelSoup mainly comes from the proposed surface-based representation and stochastic structure modeling, rather than from increased model size.
\begin{table}[hbp]
\centering
\caption{Comparison of Method Complexities.}
\label{tab:complexity_comparison_transposed}
\setlength{\tabcolsep}{3pt}
\renewcommand{\arraystretch}{0.7} % Reduces vertical spacing
\footnotesize
\begin{tabular}{l|rrrrrr}
\toprule
\multirow{2}{*}{Metric} & \multicolumn{6}{c}{Method} \\
\cmidrule{2-7}
 & SurfelSoup & Unicorn & Unicorn (large) & SparsePCGC & PCGCv2 & TriSoup \\

\midrule
Enc Time (s) & 1.1 & 1.1 & 1.1& 1.2 & 0.4 & 2.8 \\
Dec Time (s) & 1.4 & 1.2 & 1.4& 1.3 & 0.5 & 1.1 \\
Model Size (MB) & 33 & 22 & 49 & 12 & 4 & 0 \\
\bottomrule
\end{tabular}
\end{table}

\subsection{Limitations and Future Work}\label{appendix:futurework}
\paragraph{Scene Point Clouds:}
SurfelSoup shows significant gain against voxel-based baselines on dense point clouds with smooth surface structures. However, the gain on structurally complex scenes is limited because most surface areas need to be divided to the finest layer $l=1$. The potential improvements include:
\begin{itemize}
    \item {\bf Coordinate system:} SCP~\citep{luo2024scp} provides one way of preprocessing sparse LiDAR point clouds by transfering the xyz coordinate system into spherical coordinate system for more regularized point clouds. Experimental result shows around 30\% gain by simply changing the coordinate system. SurfelSoup can follow the same pipeline to make the scene point cloud denser, and more surface-like.

    \item {\bf Dataset} SurfelSoup actually has the potential of dealing with sparse point clouds due to the flexibility of pSurfels. We have mentioned in Sec.~\ref{chap:surfel_recon} that when the variance $\sigma\in\mathbb{R}^3$ is large in two axis and small in the other one, the produced pSurfel approximates a surface structure. On the other hand, when $\sigma$ is small in all the three axis, the produced pSurfel can represent tiny and irregular geometry structures, which is widely observed in sparse point clouds. Under extreme case when $\sigma\rightarrow0$, pSurfel represents a single point defined by the center $\mu$, which can perfectly depict the isolated points in sparse point clouds. Therefore, SurfelSoup may be greatly improved on sparse point clouds, if it can be trained on sparse datasets like SemanticKITTI.
\end{itemize}

\paragraph{Color:} Although SurfelSoup can be easily integrated with any point cloud attribute (color) compression codec (by recoloring the geometry, as every codec does), SurfelSoup currently does  not support attribute (color) compression. However, SurfelSoup can be extended to include attribute compression by 1) adding a texture map representation for each surfel as shown in Appendix~\ref{appendix:rendering} and compressing the texture maps; or 2) adding a separate color attribute ${\bf c}\in\mathbb{R}^3$ to pSurfel parameters. Note that a flat surface may have a complex texture pattern. Therefore, we may need multiple single-colored surfels to represent a flat surface area.
%colored point clouds have more complex surface structures compared with geometry-only point clouds ({\it e.g.}, a T-shirt's geometry can be represented by a small number of surfaces, but the pattern on that can be arbitrarily complex). Therefore, each octree node may need multiple pSurfels.

\paragraph{Model Size:} SurfelSoup's model size is 33MB. Although this is substantially smaller than widely-used vision backbones such as VGG-16 ($\sim$528 MB) or ResNet-50 ($\sim$98 MB), the model size can be further reduced by model pruning, model distillation and model quantization. Classic ``Deep Compression''~\citep{han2016deep} pipelines have shown up to 49× reduction without sacrificing accuracy on vision backbones. Recent learned image compression models demonstrate that distillation~\citep{chen2025knowledge} can reduce parameters by about 40\%.

\section{Technical Clarification Note}
\paragraph{Intuition behind pSurfel as Primitive:} A pSurfel models a local point set as a small surface patch parameterized by a generalized Gaussian (GG) function. The key intuitions are as follows:
\begin{itemize}
    \item Surface (surfel) is a more compact representation than voxels/points, when a scene contains mainly smooth surfaces and the points in each local are can be approximated by a small planar surface.  %which can represent a local point set lying on the same underlying surface. 
    This leads to both structural compactness and reconstruction smoothness.
    \item An explicit surfel representation (e.g. using a normal vector, a center position, and a radius in the tangent plane) is hard to optimize. Naively using the point-to-plane distance (D2) can be ambiguous as it restricts nothing about surfel  size. Optimizing the surfel parameters using the point-to-plane distance is also complicated and difficult to converge. Therefore, we turn to a probabilistic representation, with a continuous 3D GG function  depicting the occupancy probability of each voxel. This enables us to simply use the binary cross entropy to measure the distortion, facilitating gradient backpropagation during training. Recall that when one of the variances in the GG function is very small, it effectively represents a thin plane, and the variances in the other two axes describe the size of the plane. The ``soft'' nature of this representation allows more efficient gradient passing during training. It is also more flexible and can represent points clustered in a small area. 
\end{itemize}

\section{Reproducibility Statement}
\paragraph{Code Release:} We plan to publicly release the full implementation, including training and evaluation scripts and model weights upon acceptance of this paper.

\paragraph{Datasets:} Our experiments use publicly available datasets under MPEG Common Test Condition (CTC). Note that RWTT~\citep{maggiordomo2020real} is a mesh dataset, where we use the script from MPEG for point cloud sampling and quantization. 

\paragraph{Compute Resource:} The training is done on single NVIDIA A100 GPU for one day, identical to the training time of baselines. The peak memory consumption during training is 23 GB.

\end{document}